\documentclass[aps,pre, reprint, showpacs,superscriptaddress]{revtex4-1} 

\usepackage{latexsym}
\usepackage{amsmath,amsfonts,amssymb,amsbsy,amscd}
\usepackage[usenames,dvipsnames]{color}
\usepackage{hyperref}
\usepackage{graphicx}

\hypersetup{
   pdfauthor=Siminos,
   pdftitle=Modelling relativistic soliton interactions
}

\newcommand{\beq}{\begin{equation}}
\newcommand{\eeq}{\end{equation}}
\newcommand{\bseq}{\begin{subequations}}
\newcommand{\eseq}{\end{subequations}}

\newcommand{\rf}     [1] {~\cite{#1}}
\newcommand{\refref} [1] {Ref.~\cite{#1}}
\newcommand{\refrefs}[1] {Refs.~\cite{#1}}
\newcommand{\refeq}  [1] {Eq.~(\ref{#1})}
\newcommand{\refneq}  [1] {(\ref{#1})}
\newcommand{\refEq}  [1] {Equation~(\ref{#1})}
\newcommand{\refeqs} [2]{Eqs.~(\ref{#1})--(\ref{#2})}
\newcommand{\refEqs} [2]{Equations~(\ref{#1})--(\ref{#2})}
\newcommand{\refeqset} [2]{Eqs.~(\ref{#1}) and (\ref{#2})}
\newcommand{\reffig} [1] {Fig.~\ref{#1}}

\newcommand{\refFig} [1] {Figure~\ref{#1}}

\newcommand{\reftab} [1] {Table~\ref{#1}}

\newcommand{\refsect}[1] {Sec.~\ref{#1}}
\newcommand{\refappe}[1] {Appendix~\ref{#1}}

\newcommand{\etc}{{etc.}}       
\newcommand{\etal}{~{\em et al.}}    
\newcommand{\ie}{{i.e.}}        
\newcommand{\eg}{{e.g.}}        %

\newcommand{\Atr}{\ensuremath{\mathbf{A}_{\perp}}}
\newcommand{\Etr}{\ensuremath{\mathbf{E}_{\perp}}}
\newcommand{\sumzinf}{\sum_{j=0}^\infty}
\newcommand{\sumoneinf}{\sum_{j=1}^\infty}
\newcommand{\epsj}{\epsilon^{j}}
\newcommand{\DX}[1]{\nabla_{#1}}
\newcommand{\DT}[1]{\partial_{#1}}
\newcommand{\sech}{\mathrm{sech}}

\newcommand{\nicetilde}{\raise.17ex\hbox{$\scriptstyle\sim$}}

\newcommand{\edit}[1]{{#1}}

\newcommand{\nlse}{NLS equation}
\newcommand{\nls}{NLS}
\newcommand{\enveq}{pNLS equation}
\newcommand{\pNLS}{pNLS}
\newcommand{\swave}{soliton}
\newcommand{\swaves}{solitons}

\newcommand{\conline}{(color online)}

\newcommand{\cwidth}{246.0pt}   
\definecolor{hreflinkcolor}{rgb}{0.13,0.17,0.83}
\hypersetup{colorlinks=true,urlcolor=hreflinkcolor,
        linkcolor=hreflinkcolor,citecolor=hreflinkcolor}

\begin{document}

\author{E. Siminos}
\email{evangelos.siminos@gmail.com}
\affiliation{Max Planck Institute for the Physics of Complex Systems, N\"{o}thnitzer Str. 38,
D-01187 Dresden, Germany}

\author{G. S\'anchez-Arriaga}
\affiliation{Departamento de F\'isica Aplicada, Escuela T\'ecnica Superior
de Ingenieros Aeron\'auticos, Universidad Polit\'ecnica de Madrid, Madrid, Spain}

\author{V. Saxena}
\affiliation{Centre for Free-Electron Laser Science, Deutsches Elektronen-Synchrotron,
Notkestrasse 85, 22607 Hamburg, Germany}

\author{I. Kourakis}
\affiliation{Centre for Plasma Physics, School of Mathematics and Physics,
Queen's University Belfast, Belfast BT7 1NN, Northern Ireland, United Kingdom}

\title{Modelling relativistic soliton interactions in over-dense plasmas: a perturbed nonlinear Schr\"{o}ndinger equation framework}

\begin{abstract}
We investigate the dynamics of localized solutions of the relativistic cold fluid plasma model in the small but finite amplitude limit, for slightly
overcritical plasma density. Adopting a multiple scale analysis, we derive a perturbed nonlinear Schr\"{o}ndinger
equation that describes the evolution of the envelope of circularly polarized electromagnetic field. Retaining
terms up to fifth order in the small perturbation parameter, we derive a self-consistent framework for the description
of the plasma response in the presence of localized electromagnetic field. The formalism is applied to standing
electromagnetic soliton interactions and the results are validated by simulations of the full cold-fluid model.
To lowest order, a cubic nonlinear  Schr\"{o}ndinger equation with a focusing nonlinearity is recovered.
Classical quasiparticle theory is used to obtain analytical estimates for the collision time and minimum distance
of approach between solitons. For larger soliton amplitudes the inclusion of the fifth order terms
is essential for a qualitatively correct description of soliton interactions. The defocusing quintic nonlinearity
leads to inelastic soliton collisions, while bound states of solitons do not persist under 
perturbations in the initial phase or amplitude.
\end{abstract}

\date{\today}

\pacs{52.27.Ny, 52.35.Sb, 52.38.-r, 52.65.-y}
\keywords{laser-plasmas, relativistic solitons, solitary wave interactions, perturbed NLS}

\maketitle

\section{Introduction}

The availability of short, intense laser pulses has opened new regimes of laser-plasma interaction,
rendering possible the excitation and study of various localized structures in the plasma.
Amidst a plethora of excitations observed, particular role is played by
electromagnetic \swaves, \ie, self-trapped pulses characterized (and sustained)
by a balance of dispersion or diffraction and nonlinearity. In particular, we shall
be interested in so-called relativistic \swaves, for which the electromagnetic field
amplitude is intense enough to set plasma electrons in relativistic motion.

Relativistic \swaves\ have been predicted by analytical
theory\rf{akhiezer56,marburger1975, kozlov79,kaw92,esirkepov1998} and simulations\rf{Bulanov_92,bulanov1999,naumova2001,Esirkepov_02,bulanov2006,wu2013}, 
and their signatures have been observed in experiments\rf{borghesi2002,romagnani2010,sarri2010,sarri2011,Chen_07,Pirozhkov_07,Sylla_12}.
They can be thought of as
electromagnetic pulses trapped in a plasma density cavitation with
over-dense boundaries. For one-dimensional (1D) plasma geometry, a
vast number of \swave\ families have been identified and
studied\rf{esirkepov1998,Farina01a,Farina05,Farina05,Sanchez_2011a,Sanchez_2011b}, 
while higher-dimensional \swaves\ have also been
encountered in simulations\rf{bulanov2001b,Esirkepov_02}.
In over-dense plasmas near the critical
density, \swaves\ can be excited by a long intense
pulse incident on a plasma density gradient\rf{marburger1975,wu2013}.
In under-dense plasmas, relativistic \swaves\ have been
observed behind the wake left by an intense and short
pulse\rf{bulanov2001b,naumova2001,Esirkepov_02, bulanov2006}.
\edit{In this case, \swave\ creation 
is linked to the downshift of the laser pulse 
frequency as it propagates through the plasma, which leads
to an effective reduction of the critical density and trapping
of laser pulse energy in the form of a \swave\rf{bulanov2001b}.
Therefore, from this point on, we will use the term over-dense plasma 
in association to the frequency $\omega$ of the \swave.
The latter may be lower than the frequency of the laser pulse
that excited the \swave. 
} 

\edit{Simulations\rf{bulanov1999,bulanov2001b,wu2013} and experiments\rf{borghesi2002,romagnani2010}
show} that multiple \swaves\ may be excited by a single laser pulse, 
and one may expect that these \swaves\ can interact with each other\rf{bulanov2001b}. 
In a previous publication\rf{saxena2013} we presented numerical studies of
interactions of standing electromagnetic \swaves\ of the form
predicted in \refref{esirkepov1998}, within the
relativistic cold fluid framework. For low \swave\ amplitudes,
a phenomenology similar to Nonlinear Scr\"{o}dinger (\nlse) equation
soliton interaction\rf{hasegawa1995} has been observed, \eg\
involving the formation of bound states, under certain
circumstances. However, \swave\ interaction for larger
amplitudes departs from \nlse\ phenomenology.

The aim of the present work is to gain more insight in the origin of
the \nlse\ behavior of small amplitude \swave\ interaction in
over-dense plasmas, but also to explore how deviations from \nlse\
behavior arise. Starting from the one-dimensional relativistic cold fluid model we
develop a perturbative treatment based on multiple scale
analysis\rf{taniuti1969,dauxois2006, hoyle2006}, in which the small
parameter is the electromagnetic field amplitude. {Under the
assumptions of immobile ions and circular polarization} we derive a
perturbed \nls-type equation (\enveq) which describes the evolution
of the electromagnetic field envelope.
In our expansion, localization of the \swave\ solution is introduced naturally as a result of
the assumption of {a carrier frequency $\omega$} smaller but similar to the plasma frequency, 
{$\omega\lesssim\omega_{pe}=\sqrt{N_{0}e^2/m_e \epsilon_0}$, where $N_0$ is the plasma background density}.
The dominant nonlinear term is a `focusing' cubic nonlinearity, while at higher order a `defocusing'
quintic one also appears.
They both result from the perturbative expansion of the relativistic $\gamma$ factor.
Additional higher order nonlinear terms result from the ponderomotive coupling of the field to
the plasma. To lowest order our \enveq\ equation reduces to a focusing \nlse.

\edit{Our study of \swave\ interactions using the fluid and \enveq\ models
shows that soliton collisions are inelastic. This is commonly regarded as a signature of 
non-integrability of the governing equations\rf{fordy1990}, and is in stark contrast with the 
elastic collisions of solitons of the (integrable) cubic \nls\ equation.
We note that although a distinction between solitary waves and solitons
based on the nature of their interactions can be made\rf{scott1973},
here we adhere to the common practice in laser-plasma
interaction literature of referring to any solitary wave solution
as a soliton.}

The \nlse\ is ubiquitous in physics since it
appears generically as an envelope equation describing
propagation of weakly nonlinear waves in dispersive media\rf{hoyle2006,dauxois2006}.
It has been widely used in describing phenomena such as Benjamin-Feir type modulational instabilities,
solitons\rf{dauxois2006} and, more recently, rogue waves\rf{dysthe1999,veldes2013}.
Interestingly, the \nlse\ can be derived by symmetry considerations
alone\rf{hoyle2006}. However, for specific applications the
coefficients of the various terms must be determined through a
multiple scale analysis procedure. Then, the \nlse\ is obtained as a
compatibility condition, imposed for secular term suppression, in
third order in the small expansion parameter. The
analogous first- and second-order compatibility conditions are also
physically meaningful, as they yield the linear dispersion relation
and the associated group velocity for the envelope. Higher order
compatibility conditions contribute additional nonlinear
terms\rf{newell1992}, which may lead to a non-integrable
perturbed \nlse. In plasma physics, use of the cubic \nlse\ has a
long history, for example to describe modulated electrostatic
wavepackets\rf{kourakis2005} {and weakly relativistic
laser-plasma interactions\rf{kates1989,kuehl1993,kuehl1993b}.
For the case of linear polarization, in the highly under-dense plasma
limit $\omega_{pe}/\omega\ll1$, fifth order terms in the multiple scale 
expansion have been partially retained in \refref{kuehl1993}.
The main contribution of the present paper is the derivation of
the fifth order terms for the over-dense, near-critical case, with circular
polarization.
}

This paper is structured as follows: \refsect{s:model} recalls the relativistic cold fluid model
that will be the starting point for this work. The multiple scale expansion is described in
\refsect{s:expansion}  leading to the derivation of the \enveq, which is our main
result. In \refsect{s:nlse} we study in some detail the  \nlse\ limit of our expansion, comparing
numerical results of cold fluid model simulations for \swave\ interaction to classical predictions
of quasiparticle theory for soliton interaction. In \refsect{s:numerics} we compare numerical
simulations of \swave\ interaction in
the three levels of description: cold fluid model, \enveq\ and \nlse. In \refsect{s:concl} we discuss
our findings and present our conclusions. \edit{\refappe{s:ic} describes the relation between fluid
and envelope initial conditions, while \refappe{s:numfluid} provides details on the numerical implementation
of the fluid code.}

\section{\label{s:model} Relativistic cold fluid model}

Our starting point is the relativistic cold fluid plasma-Maxwell model (see, for example,
\refref{gibbon2005} for the derivation and history of the model) in one spatial dimension.
We assume a cold plasma of infinite extent with immobile ions that provide
a neutralizing background.
Considering infinite plane waves, propagating along the $x$-direction
and working in the Coulomb gauge, the longitudinal and transverse
component of Ampere's law are expressed as
\begin{subequations}\label{eq:fluid}
  \begin{equation}\label{eq:longitudinal}
    \frac{\partial^2 \Phi}{\partial x\partial t}+\frac{N}{\gamma}P_x=0\,,
  \end{equation}
 and
  \begin{equation}\label{eq:wave}
    \frac{\partial^2\textbf{\Atr}}{\partial x^2}-\frac{\partial^2\textbf{\Atr}}{\partial t^2}=\frac{N}{\gamma}\textbf{\Atr}\,,
  \end{equation}
respectively. The fluid momentum equation is
  \begin{equation}\label{eq:Px}
    \frac{\partial P_x}{\partial t}=\frac{\partial}{\partial x}\left(\Phi-\gamma\right)\,,
  \end{equation}
while Poisson equation yields
  \begin{equation}\label{eq:Poisson}
    N=1+\frac{\partial^2\Phi}{\partial x^2}\,.
  \end{equation}
Here,
  \begin{equation}\label{eq:gamma}
    \gamma=\sqrt{1+P_x^2+\Atr^2}\,,
  \end{equation}
\end{subequations}
is the relativistic factor, while
$\mathbf{P}(x,t)=\mathbf{p}(x,t)-\mathbf{A}(x,t)$ is the generalized
momentum of the electron fluid, $\mathbf{p}(x,t)$ the kinetic
momentum normalized to $m_ec$, $N$ is the fluid charge density
normalized to the plasma background density $N_0$,
$\mathbf{\Atr}=A_y\mathbf{\hat{j}}+A_z\mathbf{\hat{k}}$ is
the transverse vector potential and $\Phi$ is the scalar
potential, both normalized to $m_ec^2/e$. Moreover, time
and length are respectively normalized to the inverse of the plasma
frequency $\omega_{p0}$ and the
corresponding skin depth $c/\omega_{p0}$.
In our one-dimensional
modelling we have taken the longitudinal vector potential $A_x=0$, while we have used
the conservation of transverse canonical momentum, assuming an initially cold plasma, to write
\refeq{eq:wave} and \refeq{eq:gamma}.

\section{\label{s:expansion} Multiple scale expansion}

Multiple scale analysis (see, e.g. \refref{dauxois2006}) seeks to
describe perturbatively a system of differential equations by
assuming it evolves in different, well separated temporal and
spatial scales. Our treatment follows the standard procedure of
removal of secular terms by imposing suitable solvability conditions
at each order in perturbation analysis. However, our treatment of
the dispersion relation in \refsect{s:dispersion} is novel and has
been devised in order to relate the electromagnetic field
frequency to the small parameter of the expansion, while introducing
localization of solutions in an over-dense plasma in a natural way.

We proceed
by introducing the scaled (or slow) variables $T_j = \epsj t$ and $X_j = \epsj x$, where $\epsilon$ is a small
parameter and $j=0,1,2,\ldots\,$. We assume that the fields within the plasma are small,
so that we may expand the fluid model variables as
\begin{subequations}\label{eq:exp}
  \begin{equation}\label{eq:expAy}
   A_y (x,t) = \sumoneinf \epsj a_j(X_0,X_1,\ldots,T_0,T_1,\ldots)\,,
  \end{equation}
  \begin{equation}\label{eq:expAz}
   A_z (x,t) = \sumoneinf \epsj b_j(X_0,X_1,\ldots,T_0,T_1,\ldots)\,,
  \end{equation}
  \begin{equation}\label{eq:expPhi}
   \Phi (x,t) = \sumoneinf \epsj \phi_j(X_1,\ldots,T_1,\ldots)\,,
  \end{equation}
  \begin{equation}
   P_x (x,t) =  \sumoneinf \epsj p_j(X_1,\ldots,T_1,\ldots)\,,
  \end{equation}
  \begin{equation}\label{eq:Nexp}
   N (x,t) =  \sumzinf \epsj n_j(X_1,\ldots,T_1,\ldots)\,,
  \end{equation}
  and
  \begin{equation}\label{eq:gammaExp}
   \gamma = \sumzinf \epsj \gamma_j =  1 + \frac{1}{2}(p_1^2+a_1^2+b_1^2)\epsilon^2+\ldots\,.
  \end{equation}
  Moreover we will need
  \begin{equation}\label{eq:deltaExp}
   \delta \equiv 1/\gamma = \sumzinf \epsj \delta_j =  1 - \frac{1}{2}(p_1^2+a_1^2+b_1^2)\epsilon^2+\ldots\,.
  \end{equation}
\end{subequations}

In writing \refeq{eq:Nexp} we took into account \refeqset{eq:Poisson}{eq:expPhi}.
We assumed that the longitudinal quantities $\Phi,\,P_x,\, N$ do not depend on the
fast time scale $T_0$ associated with the transverse electromagnetic field oscillations,
{and took into account that for CP pulses there is no generation of harmonics of the basic frequency
$\omega$\rf{kates1989,kaw92}}.
These assumptions are consistent with CP \swave\ solutions of \refref{esirkepov1998}, the
interactions of which we shall study here. 

Defining $\DT{i}\equiv \partial/\partial T_i$ and $\DX{i}\equiv \partial/\partial X_i$,
we get
\begin{subequations}\label{eq:der}
  \beq
    \frac{\partial}{\partial t} = \DT{0} + \epsilon \DT{1} + \epsilon^2 \DT{2} + \ldots\,,
  \eeq
  \beq
    \frac{\partial}{\partial x} = \DX{0} + \epsilon \DX{1} + \epsilon^2 \DX{2} + \ldots\,,
  \eeq

  \beq
    \frac{\partial^2}{\partial x \partial t} = \DX{0}\DT{0} + \epsilon (\DX{0}\DT{1} + \DX{1}\DT{0}) + \ldots\,,
  \eeq
\end{subequations}
\etc

We proceed by substituting \refeq{eq:exp} and \refeq{eq:der} into \refeq{eq:fluid}, and collecting
terms in different orders of $\epsilon$.

\subsection{Order \texorpdfstring{$\epsilon^0$}{0}}

The only equation that contains terms of order $\epsilon^0$ is \refeq{eq:Poisson}, which gives
\beq\label{eq:n0}
  n_0(X_0,X_1,\ldots,T_0,T_1,\ldots) = 1\,.
\eeq

\subsection{\label{s:dispersion} Order \texorpdfstring{$\epsilon^1$}{1}: linear dispersion relation}

Collecting terms of order $\epsilon$, \refeqs{eq:longitudinal}{eq:Px} give
$p_1  = n_1 = 0$, while $\phi_1$ remains unspecified.
Taking into account \refeq{eq:n0}, we get for the $y$ component of \refeq{eq:wave},
\beq\label{eq:wave1}
  \DX{0}^2 a_1 - \DT{0}^2 a_1 = a_1\,.
\eeq
We are interested in plane wave solutions of this linear wave equation, which have the form:
\begin{equation}\label{eq:a1}
  a_1 (X_0,X_1,\ldots,T_0,T_1,\ldots)=  
          a(X_1,\ldots,T_1,\ldots) e^{i(k X_0 - \omega T_0)}\, +\, c.c.\,,
\end{equation}
where $+\, c.c.$ denotes the complex conjugate of the preceding expression.
Plugging \refeq{eq:a1} back into \refeq{eq:wave1}, we see that these solutions satisfy the
usual plasma dispersion relation
\beq\label{eq:disp}
  \omega^2 = 1 + k^2\,.
\eeq
Here, we are interested in over-dense plasmas, in which $\omega<1$ and thus
\beq
  k^2=\omega^2-1<0\,,
\eeq
\ie, $k$ is imaginary and the waves are localized. Moreover, we are interested here in
describing solutions close to small amplitude \swaves,
which have $\omega\simeq1$. Therefore,  $|k|$ is small and this fact in combination with
the dispersion relation, \refeq{eq:disp}, suggests the following expansions
\beq\label{eq:kwExp}
    \omega = \sumzinf \epsj \omega_j\,,\qquad k = \sumoneinf \epsj k_j\,.
\eeq

Substituting \refeq{eq:kwExp} into the dispersion relation \refeq{eq:disp} we obtain
\beq\label{eq:dispAppr}
  \omega = 1 - \frac{\epsilon^2}{2}|k_1|^2+\ldots\,,
\eeq
where we used $k_1=\pm i|k_1|$. This implies
that expressions such as $\epsilon|k_1|\,X_0$, $\epsilon^2|k_1|^2 T_0$ \etc\
appearing in the oscillating part of \refeq{eq:a1}, do in fact represent
slow variations. Thus, we may include them into the envelope, \ie, we may
write \refeq{eq:a1} as
\begin{equation}\label{eq:a0corr}
  a_1 (X_0,X_1,\ldots,T_0,T_1,\ldots) =\\ a(X_1,\ldots,T_1,\ldots) e^{-i T_0}\,+\,c.c.\,.
\end{equation}
Note that the dependence on $X_0$ has naturally dropped.

\subsection{Order \texorpdfstring{$\epsilon^2$}{2}}

Collecting terms of order $\epsilon^2$ we get from \refeq{eq:longitudinal} and \refeq{eq:Poisson}
$p_2  = n_2 = 0$, while \refeq{eq:Px} yields
\beq\label{eq:Px2}
  \DX{1}\phi_1-\frac{1}{2}\DX{0}(a_1^2+b_1^2) = 0 \,,
\eeq
respectively. For circular polarization (CP), we have $a= - i\,b$, and therefore \refeq{eq:gammaExp}
yields
\beq
  \gamma_2 = \frac{1}{2}(a_1^2+b_1^2)=2\,|a|^2\,,
\eeq
independent of $T_0$. Thus, \refeq{eq:Px2} simplifies to
\beq\label{eq:DX1phi1}
  \DX{1}\phi_1=0\,.
\eeq

Collecting terms of order $\epsilon^2$ we get for the $y$ component
of \refeq{eq:wave}, 
\begin{equation}\label{eq:wave2}
  \mathcal{L} a_2  = 2 \DT{0} \DT{1} a_1  = -2\,i\, \frac{\partial a}{\partial T_1} e^{-i T_0}\,,
\end{equation}
where we introduced the operator
\begin{equation}
  \mathcal{L} \equiv (\DX{0}^2-\DT{0}^2-1)\,.
\end{equation}
 The term on the right hand side is resonant forcing term for the
linear operator $\mathcal{L}$. Therefore we impose the solvability
condition \beq\label{eq:solv2}
    \frac{\partial a}{\partial T_1} = 0\,.
\eeq
Then \refeq{eq:wave2} becomes $\mathcal{L}a_2 = 0\,,$.
It has plane wave solutions that can be included in $a_1$ in \refeq{eq:a0corr}, 
so we may take $a_2 = 0$. From \refeq{eq:gammaExp} we then find $\gamma_3 =0$.

\subsection{Order \texorpdfstring{$\epsilon^3$}{3}: \nlse\ limit}

Collecting terms of order $\epsilon^2$ and using \refeq{eq:DX1phi1},
we get from \refeqs{eq:longitudinal}{eq:Px}, $p_3=n_3=0$,
and
\beq\label{eq:Px3}
  \DX{2}\phi_1+\DX{1}(\phi_2-2|a|^2) = 0\,,
\eeq
respectively. Operating on \refeq{eq:Px3} with $\DX{1}$ and using
\refeq{eq:DX1phi1}, we obtain
\beq\label{eq:Px3poisson}
  \DX{1}^2(\phi_2-2|a|^2) = 0\,,
\eeq
which, requiring that $\phi_2$ and $a$ vanish as $X_1\rightarrow\infty$, implies
\beq\label{eq:phi2a}
  \DX{1}(\phi_2 -2|a|^2) = 0\,.
\eeq
Then in turn, \refeq{eq:Px3} implies
\beq\label{eq:DX2phi1}
  \DX{2}\phi_1 = 0\,.
\eeq

Collecting terms of order $\epsilon^2$ we get for the $y$ component of \refeq{eq:wave},
\begin{align}\label{eq:wave3}
  \mathcal{L} a_3 & = -\DX{1}^2 a_1 + (\DT{1}^2 + 2 \DT{0} \DT{2}) a_1 - 2|a|^2 a_1\,,\\
                  & = -\frac{\partial^2 a}{\partial X_1^2} e^{-i T_0} - 2\,i\, \frac{\partial a}{\partial T_2} e^{-i T_0} - 2\,|a|^2 a\, e^{-i T_0}\,,
\end{align}
where we also used \refeq{eq:solv2}.
The terms in the right hand side are all resonant with the operator
$\mathcal{L}$, leading to the solvability condition
\beq\label{eq:NLS}
  i\, \frac{\partial a}{\partial T_2} + \frac{1}{2}\frac{\partial^2 a}{\partial X_1^2} + |a|^2 a = 0\,.
\eeq
This is a {nonlinear Schr\"ondinger (\nls) equation}.
Imposing the solvability condition \refeq{eq:NLS}, we get from \refeq{eq:wave3},
$\mathcal{L}a_3 = 0$, or $a_3 =0$.

According to the standard multiple scale treatment of waves in the fluid plasma description,
the cubic \nlse~\refeq{eq:NLS} is obtained at this order, and the iterative expansion procedure stops here.
{An exception is \refref{kuehl1993}, where the limit of small density ($\omega_{pe}\ll\omega$)
has been considered and a different scaling of the slow variables has been derived, partially including terms
of fifth order in $\epsilon$.
The absence of harmonic terms for circular polarization makes the $\epsilon$-expansion much simpler 
in our case and allows us to keep all terms up to fifth order in the following, 
deriving a single perturbed NLS equation.}

\subsection{Order \texorpdfstring{$\epsilon^4$}{4}}

From \refeq{eq:longitudinal} we get, with the
help of \refeq{eq:DX1phi1} and \refeq{eq:DX2phi1},
\beq
  \DX{1}\DT{1}\phi_2 + p_4 = 0\,.
\eeq
Operating with $\DT{1}$ on \refeq{eq:phi2a} and using \refeq{eq:solv2} we obtain
\beq
  \DX{1}\DT{1}\phi_2 = 0\,,
\eeq
and therefore $p_4 = 0$.

\refEq{eq:Poisson} gives
\beq\label{eq:n4}
  n_4 = \DX{1}^2\phi_2\,.
\eeq

Noting that
\beq
  \gamma_4 = -2\,|a|^4\,,
\eeq
we get from \refeq{eq:Px}
\beq\label{eq:Px4}
  \DX{1}\phi_3 + \DX{3}\phi_1+\DX{2}(\phi_2 + 2|a|^2) =0 \,.
\eeq

Collecting terms of order $\epsilon^4$ we get for the $y$ component
of \refeq{eq:wave}, 
\begin{align}\label{eq:wave4}
  \mathcal{L} a_4 & =  2\DT{0}\DT{3}a_1 - 2 \DX{1}\DX{2}a_1\\
                  & = -2\,i\DT{3}a\, e^{-iT_0} - 2 \DX{1}\DX{2}a\, e^{-iT_0}\,.
\end{align}
Requiring that resonant forcing terms vanish, we obtain the solvability condition
\beq\label{eq:solv4}
  i\DT{3}a + \DX{1}\DX{2}a =0\,.
\eeq
Finally, \refeq{eq:wave4} implies $a_4$ may be included into $a_1$ and we write $a_4 = 0$.

\subsection{Order \texorpdfstring{$\epsilon^5$}{5}}

\refEq{eq:longitudinal} gives
\beq\label{eq:longitudinal5}
  p_5 + \DX{3}\DT{1}\phi_1 + \DX{2}\DT{1}\phi_2+\DX{1}\DT{2}\phi_2+\DX{1}\DT{1}\phi_3=0\,,
\eeq
where we have used \refeq{eq:DX2phi1} and \refeq{eq:DX1phi1}. Applying $\DX{1}$ on \refeq{eq:Px4}
and using \refeq{eq:DX1phi1} and \refeq{eq:phi2a} we get
\beq\label{eq:phi3Lapl}
  \DX{1}^2\phi_3 = 0\,,
\eeq
which implies
\beq\label{eq:DX1phi3}
  \DX{1}\phi_3 = 0\,.
\eeq
On the other hand, applying $\DT{1}$ on \refeq{eq:Px4} and using \refeq{eq:solv2} and \refeq{eq:DX1phi3}
we get
\beq\label{eq:DX3DT1phi1}
  \DX{3}\DT{1}\phi_1+\DX{2}\DT{1}\phi_2=0\,.
\eeq
Thus, we may use \refeqs{eq:DX1phi3}{eq:DX3DT1phi1} to get from \refeq{eq:longitudinal5}
\beq\label{eq:p5}
  p_5 = -\DX{1}\DT{2}\phi_2 = -2\DT{2}\DX{1}|a|^2\,,
\eeq
where, in the last step, we have used \refeq{eq:phi2a}.

From Poisson equation \refeq{eq:Poisson} we obtain
\beq\label{eq:n5}
  n_5 = 2 \DX{2}\DX{1}\phi_2 = 4\DX{2}\DX{1}|a|^2\,.
\eeq

\refEq{eq:Px} gives
\beq\label{eq:Px5}
  \DX{1}(\phi_4-\gamma_4)+\DX{2}\phi_3+\DX{3}(\phi_2-\gamma_2)+\DX{4}\phi_1 = 0\,.
\eeq
Operating on \refeq{eq:Px5} with $\DX{1}$ and using \refeq{eq:phi2a}, \refeq{eq:Px3poisson} and \refeq{eq:DX1phi3} we
obtain
\beq\label{eq:Px5poisson}
  \DX{1}^2(\phi_4-\gamma_4) = 0\,.
\eeq

Noting that $\delta_4=6\,|a|^4$, the wave equation gives at order $\epsilon^5$
\beq\label{eq:wave5}
  \mathcal{L}\, a_5 = -2\DX{1}\DX{3}a_1-\DX{2}^2 a_1 +2 \DT{0}\DT{4}a_1+\DT{2}^2 a_1 + 6 |a|^4 a_1 + n_4 a_1\,.
\eeq
All terms on the right hand side are resonant forcing terms, and with the help of \refeq{eq:n4}
and \refeq{eq:Px3poisson} we are led to the solvability condition
\beq\label{eq:solv5}
  -2\DX{1}\DX{3}a-\DX{2}^2 a -2\,i\,\DT{4}a  +\DT{2}^2 a + 6 |a|^4 a + 2 a\DX{1}^2|a|^2  = 0\,.
\eeq
We note, that we do not employ a parabolic approximation but rather maintain the term involving second time
derivative. Finally, \refeq{eq:wave5}, allows one to write $a_5 = 0$.

\subsection{Collecting different orders: \enveq}

The strategy we follow in order to write a perturbed \nlse, inspired by \refref{newell1992} (Chapter 3),
is to multiply the solvability conditions
imposed at various orders with appropriate constants, so as to form the expansions of the differential operators
\refeq{eq:der}. Specifically, we form the sum
\beq
  2\,i\,\epsilon\times(\ref{eq:solv2}) + 2\,\epsilon^2\times(\ref{eq:NLS}) + 2\,\epsilon^3\times(\ref{eq:solv4})- \epsilon^4\times(\ref{eq:solv5})\,,
\eeq
which, using \refeq{eq:der}, reads
\beq
  i\frac{\partial a}{\partial t} + \frac{1}{2}\frac{\partial^2 a}{\partial x^2}+\epsilon^2 |a|a - 3\,\epsilon^4 |a|^4 a -\epsilon^2 \frac{\partial^2 |a|^2}{\partial x^2} a - \frac{1}{2} \frac{\partial^2 a}{\partial t^2}= 0
\eeq

Rescalling,
\beq\label{eq:scale}
 X=\epsilon x,\qquad T=\epsilon^2 t\,,
\eeq
we arrive at
\beq\label{eq:env}
  i\frac{\partial a}{\partial T}+\frac{1}{2}\frac{\partial^2 a}{\partial X^2}+|a|^2 a = \epsilon^2\left( 3\, |a|^4 a + \frac{\partial^2|a|^2}{\partial X^2} a + \frac{1}{2}\frac{\partial^2 a}{\partial T^2}\right)\,,
\eeq
which is our main result. \refEq{eq:env} has the form of a singularly perturbed nonlinear
Schr\"{o}dinger equation (\enveq), since the highest order time derivative
appears in the perturbation term. Note that, due to our treatment of the dispersion
relation in \refsect{s:dispersion}, \refeq{eq:env} is written in the lab frame,
without the need of a coordinate transformation to a frame moving with the
group velocity $v_g=\omega'(k)$ which is usually required in the derivation
of \nls-type equations. We therefore bypass any problems related to
the fact that $\omega'(k)$ as obtained from \refeq{eq:disp} is imaginary.

Solution of \refeq{eq:env} for $a(X,T)$ also determines the rest of the variables
in the perturbation expansion of the cold fluid model. Indeed, we
get from \refeq{eq:Nexp}, \refneq{eq:n0}, \refneq{eq:n4} and \refneq{eq:n5},
in a similar manner as above,
\beq\label{eq:n}
  N = 1+ 2\,\epsilon^4\frac{\partial^2 |a|^2}{\partial X^2}+\mathcal{O}(\epsilon^6)\,,
\eeq
\ie, $n_5$ may be included in the fourth order terms.

Similarly, \refeq{eq:p5} gives
\beq\label{eq:p}
  P = - 2\, \epsilon^5 \frac{\partial^2 |a|^2}{\partial T \partial X} + \mathcal{O}(\epsilon^6)\,,
\eeq

Finally, noting that \refeq{eq:DX1phi1} yields
\beq\label{eq:phi1Lapl}
  \DX{1}^2\phi_1=0\,,
\eeq
and forming the sum
\beq
  \epsilon\times(\ref{eq:phi1Lapl})+\epsilon^2\times(\ref{eq:Px3poisson})+\epsilon^3\times(\ref{eq:phi3Lapl}) +\epsilon^4\times(\ref{eq:Px5poisson})\,,
\eeq
we obtain
\beq\label{eq:longBalance}
  \frac{\partial^2}{\partial X^2}(\Phi-\gamma) = \mathcal{O}(\epsilon^5)\,.
\eeq
Noting that $\Phi$, $\gamma$, do not depend on $X_0$,
and taking into account boundary conditions at $x\rightarrow\pm\infty$, we obtain
\begin{align}
  \Phi  & = \gamma-1+\mathcal{O}(\epsilon^5)\label{eq:phigamma}\\
        & = 2|a|^2\,\epsilon^2 - 2|a|^4\,\epsilon^4 +\mathcal{O}(\epsilon^5)\,.
\end{align}
Therefore, the solution of \refeq{eq:env} for $a(X,T)$, also determines $N,\, P$ and $\Phi$.

The cubic $|a|^2 a$ and quintic terms $3\,\epsilon^2 |a|^4 a$ come from the expansion of the relativistic
factor $\gamma$ and therefore correspond to relativistic corrections
to the optical properties of the plasma, \ie, they pertain to transverse dynamics.
The cubic term results in compression of the pulse, while the quintic term
has the opposite effect (due to the difference in sign). The term  $a\,\partial_{XX}|a|^2$ takes into account longitudinal effects,
namely the coupling of the electromagnetic field to the plasma through charge separation caused by
ponderomotive effects [see \refeq{eq:n}]. Moreover, \refeq{eq:p} shows that
the balance of ponderomotive $\partial_X \gamma$ and electrostatic force $\partial_X \phi$
implied by \refeq{eq:longBalance}, is violated at order $\epsilon^5$ [see also \refeq{eq:Px}].

\section{\label{s:nlse} \nlse\ limit}

Neglecting  terms of order $\epsilon^2$, \refeq{eq:env}
reduces to a \nlse\rf{sulem1999}
\beq\label{eq:NLSsimple}
  i\, \frac{\partial a}{\partial T} + \frac{1}{2}\frac{\partial^2 a}{\partial X^2} + |a|^2 a = 0\,,
\eeq
which has both moving and standing envelope soliton solutions, which we now briefly study in
order to establish connection with cold-fluid model \swaves.

\subsection{Moving soliton solutions}

Moving soliton solutions of \refeq{eq:NLSsimple} have the form\rf{dauxois2006}
\beq\label{eq:nlsMoving}
  a(X,T) = \alpha_0\, \sech\left[\alpha_0 (X-u_e\,T)\right]e^{i\,u_e\,(X-u_p\,T)}\,,
\eeq
where $u_e$ is the envelope (or group) velocity, $u_p$ is the phase velocity, and the
amplitude $a_0$ is given by
\beq
  \alpha_0=\sqrt{u_e^2-2\,u_e\,u_p}\,.
\eeq
Note that $u_e$ and $u_p$ refer to the scaled variables $X$ and $T$. We can go back to
the original variables by using \refeq{eq:AyAzEnv} of \refappe{s:ic}, which gives
\beq
  A_y =  F_0(x,t)\,\cos\left[u_e\epsilon x-(1+u_e u_p \epsilon^2)t\right]\,,
\eeq
where $F_0(x,t)=2\, \alpha\, \epsilon\,\sech\left[\alpha_0\epsilon\left(x-\epsilon u_e t\right)\right]$. 
A similar expression can be written for $A_z$, while
\refEq{eq:n}, \refneq{eq:p} and \refeq{eq:phigamma} provide
expressions for the remaining fluid variables. We see that the phase and group velocity
read
\beq
  v_{ph} = \frac{1+u_e u_p \epsilon^2}{u_e \epsilon}\,,
\eeq
and
\beq
  v_{g} = \epsilon u_e\,,
\eeq
respectively. Therefore, in terms of the cold fluid model, moving soliton solutions \refeq{eq:nlsMoving}
represent slowly propagating \swaves.

\begin{figure*}[ht!]
 \includegraphics[width=\textwidth, clip=true]{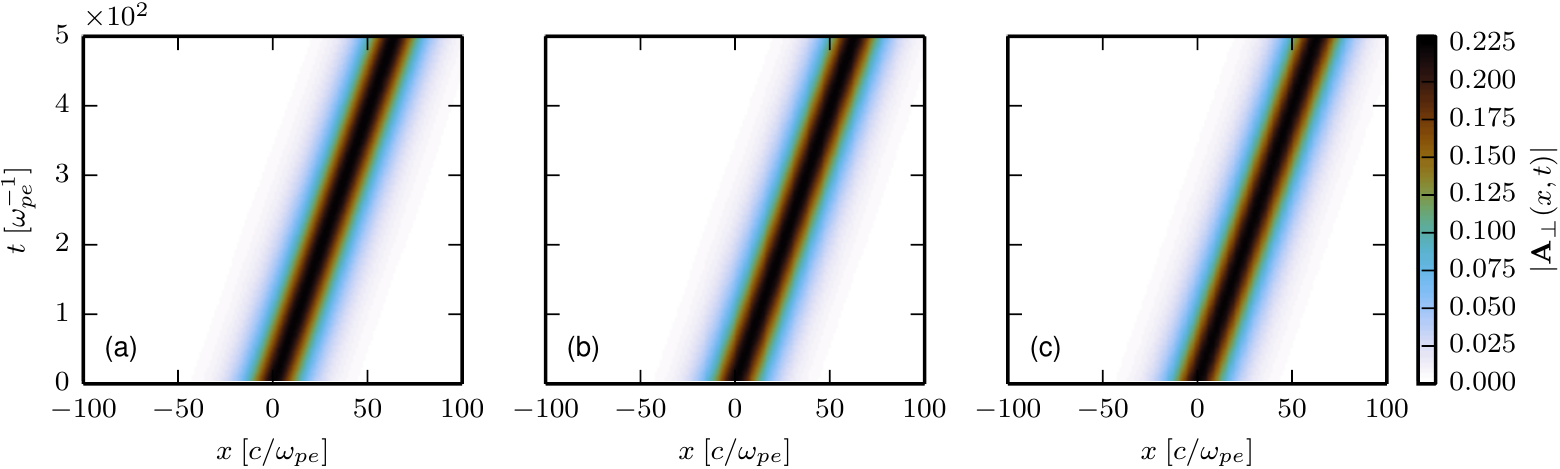}
\caption{\conline\label{fig:099_moving} Simulation of a propagating \swave\ with $u_e=0.9$, $u_p=0.1$ and $\epsilon=0.141$,
using (a) the fluid model \refeq{eq:fluid}, (b) \enveq\ \refeq{eq:env},
(c) \nlse\refneq{eq:NLS}.}
\end{figure*}

We show an example of propagating \nlse\ solitons
\refeq{eq:nlsMoving} with  $u_e=0.9$, $u_p=0.1$ and $\epsilon=0.141$
in \reffig{fig:099_moving}. Although this is only an approximate
\swave\ solution of the \enveq\ and cold fluid model, we
observe propagation at the predicted group velocity $\epsilon u_e$,
while the solution approximately maintains its shape. An exact
propagating \swave\ solution would have to be determined by
methods similar to those used, for example in \refrefs{kaw92,Sanchez_2011a}.
This is however outside the scope of this work which
emphasizes standing \swave\ interactions.

\subsection{Standing soliton solutions}

Standing soliton solutions of \refeq{eq:NLSsimple} have the form
\beq\label{eq:NLSsol}
  a(X,T) = \phi(X) e^{- i\lambda T}\,,
\eeq
where the frequency $\lambda$ is to be determined. A well known \swave\ family of \refEq{eq:NLSsimple} has
the form (see, e.g., \refref{dauxois2006})
\beq\label{eq:sech}
  \phi(X,T) = \alpha\, \sech[\alpha(X-X_0)]\,,
\eeq
where the amplitude $\alpha$ is constant. Substituting \refeqs{eq:NLSsol}{eq:sech}
into \refeq{eq:NLSsimple} we obtain
\beq\label{eq:lambda}
  \lambda=-\frac{1}{2}\alpha^2\,.
\eeq

Using \refeq{eq:AyAzEnv}  of \refappe{s:ic}, one may go back to the original
cold-fluid model variables,
\begin{subequations}\label{eq:sechA}
  \beq\label{eq:sechAy}
    A_y(x,t) = C_0(x)\,\cos(t-\epsilon^2\alpha^2\,t/2)\,,
  \eeq
  \beq
    A_z(x,t) = C_0(x)\,\sin(t-\epsilon^2\alpha^2\,t/2)\,,
  \eeq
\end{subequations}
where $C_0(x)=2\epsilon \alpha\, \sech[\epsilon\alpha(x-x_0)]$.
Note that the term $-\epsilon^2 \alpha^2\,t/2$ corresponds to the term $-\epsilon^2 |k_1|^2\,t/2$ in
\refeq{eq:dispAppr} which has been included into the envelope. This correction to $\omega$ therefore
corresponds to the slow oscillations of the soliton envelope, and we may identify $\alpha= |k_1|$.

A connection  to exact cold-fluid mode \swaves\ of Esirkepov~\etal\rf{esirkepov1998} which have
  \begin{equation}\label{eq:Esirkepov0}
  A_y= R(x)\, \cos(\omega t) \,,\qquad A_z= R(x)\, \sin(\omega t)\,,
  \end{equation}
where
\beq\label{eq:R}
  R(x) = \frac{2\sqrt{1-\omega^2}\cosh\left[\sqrt{1-\omega^2}(x-x_0)\right]}
          {\cosh^2\left[\sqrt{1-\omega^2}(x-x_0)\right]+\omega^2-1}
\eeq
can now be established. Taking into account that
\refeq{eq:dispAppr} yields 
\beq\label{eq:omegaepsk1}
  \omega^2=1-\epsilon^2\,|k_{1}|^2,
\eeq
$A_y$ from \refeq{eq:Esirkepov0} becomes, in leading order in $\epsilon$,
\begin{equation}\label{eq:EsirkepovSmall}
 A_y=
  2\epsilon|k_{1}|\sech\left[\epsilon|k_{1}|(x-x_0)\right]\cos(t-\epsilon^2\,|k_{1}|^2\,t/2)\,.
\end{equation}

Comparison of \refeq{eq:sechAy} with \refeq{eq:EsirkepovSmall}, once
again shows that we may identify $\alpha=|k_{1}|$. The value of
$|k_1|$ still remains unspecified, \edit{apart from the requirement that
$|k_1|\sim\mathcal{O}(1)$ which follows from \refeq{eq:kwExp}}. For the study of a single
soliton, or interacting \swaves\ of the same amplitude, we
may, without loss of generality, set $|k_{1}|=1$, \ie, work with
$\alpha=1$ in \refeq{eq:sechA}. Then, the small parameter $\epsilon$
in the expansion is related to $\omega$ through
\beq\label{eq:omegaeps}
  \epsilon=\sqrt{1-\omega^2}\,.
\eeq
However, the indeterminancy of $k_1$ in \refeq{eq:EsirkepovSmall} allows to model interactions of \swaves\
with different frequences (and therefore amplitudes) using a single small parameter
$\epsilon$, see \refsect{s:numerics}.

\subsection{\label{s:quasiparticle} Quasiparticle approach to soliton interactions}

Soliton interactions of \nlse\ \refneq{eq:NLSsimple} have been
studied through different methods and are well understood,
particularly when the separation of the two solitons is large, see
for example \refref{hasegawa1995}. In that case, one may consider
the solitons as 'quasiparticles', \ie, independent entities that
exert a force to one another and to a good approximation maintain
their shape during their
interaction\rf{karpman1981,gordon1983,hasegawa1995}
{(adiabatic approximation)}. One then may approximate the
evolution of the solitons with a set of few coupled ordinary
differential equations for the evolution of the soliton parameters.

A particularly useful perspective is that of Gordon\rf{gordon1983},
who shows that within this framework the dynamics of two solitons
may be described by the system of ODEs
\begin{subequations}\label{eq:gordon}
  \begin{align}
 \ddot{D} &=-8\,e^{-D}\,  \cos(\Psi)\,,\\
 \ddot{\Psi} &= 8\,e^{-D}\, \sin(\Psi)\,,
  \end{align}
\end{subequations}
where $D(T)$ and $\Psi(T)$ are the soliton peak-to-peak distance and
relative phase, respectively, $1\pm\dot{\Psi}/2$ are their
amplitudes and a dot indicates derivative with respect to $T$.
Therefore, the solitons exert to each other an effective force with
magnitude that decreases exponentially with distance $D$ and sign
that depends on their relative phase $\Psi$.

For the particular case of solitons of equal initial amplitude,
$\dot{\Psi}_0=0$, and zero initial phase difference, we see
from \refeq{eq:gordon} that the force is attractive, while the
relative phase remains zero, leading to the formation of an
oscillatory bound state. A detailed
calculation\rf{gordon1983,hasegawa1995} shows that the peak-to-peak
separation varies as
\begin{equation}\label{eq:DX}
  D(T)  = D_0 + 2\ln|\cos(2\,e^{-D_0/2} \, T)|\,,
\end{equation}
where $D_0\gg1$ is the initial distance. \refEq{eq:DX} implies that
the time after which the solitons collide for the first time is\rf{gordon1983,hasegawa1995}
\begin{equation}\label{eq:Tcoll}
  T_{coll}=\frac{1}{2}e^{D_0/2}\cos^{-1}\left(e^{-D_0/2}\right) \simeq \frac{\pi}{4}e^{D_0/2}\,.
\end{equation}
We note that in scaled variables collision time $T_{coll}$ is, according to \refeq{eq:Tcoll},
independent of $\omega$. However, when we go back to units of
$\omega_{pe}^{-1}$ and $c/\omega_{pe}$ for time and space, respectively,
through \refeqset{eq:scale}{eq:omegaeps},
we obtain
\begin{equation}\label{eq:tc}
  t_{\mathrm{coll}}=\frac{\pi}{4(1-\omega^2)}e^{\sqrt{(1-\omega^2)}\,d_0/2}\,.
\end{equation}
where $t_{\mathrm{coll}}=T_{coll}/\epsilon^2$ and $d_0=D_0/\epsilon$. For
large enough $d_0$, $t_{\mathrm{coll}}$ decreases for increasing
$\omega$.
Analytical predictions of \refeq{eq:tc} is compared with
numerical simulations in \refsect{s:numerics}.

For solitons of equal initial amplitude, $\dot{\Psi}_0=0$
and initial relative phase $\Psi_0$, \edit{the seperation versus time reads\rf{hasegawa1995}
\beq\label{eq:DwPhase}
  D(T) = D_0 + \ln\left|\frac{\cosh(-\kappa_1\,T)+\cos(\kappa_2\,T)}{2}\right|\,,
\eeq
where 
\[
  \kappa_1=4\,e^{-D_0/2}\,\sin\left(\Psi_0/2\right),\qquad  \kappa_2=4\,e^{-D_0/2}\cos\left(\Psi_0/2\right)\,. 
\]
Thus, the solitons eventually drift apart since, as we see from \refeq{eq:gordon}, the force is not 
attractive for all $T$. For large $T$, after the solitons drift apart, 
the soliton amplitudes differ by 
\beq\label{eq:Da}
    |\Delta a| = |a_1-a_2| = |4\,e^{-D_0/2}\cos\left(\Psi_0/2\right)|\,.
\eeq
}
The minimum peak-to-peak distance
is
 \beq\label{eq:Dmin}
  D_{min} = D_0+\ln\left[\frac{1}{2}\left(\cosh\left(-\pi\tan\frac{\Psi_0}{2}\right) -1\right)\right]
\eeq
reached at time
\begin{equation}
 T_{min}=\frac{\pi}{4\cos(\Psi_0/2)}e^{D_0/2}
\end{equation}
or, in units of $\omega_{pe}^{-1}$,
\begin{equation}
  t_{\mathrm{min}}=\frac{\pi\,e^{\sqrt{(1-\omega^2)}\,d_0/2}}{4(1-\omega^2)\cos(\Psi_0/2)}\,.
\end{equation}
From \refeq{eq:Dmin} one finds\rf{hasegawa1995} that the collision
is inhibited for initial phase $\Psi_0>\Psi_c$, where 
\beq
  \tan\frac{\Psi_c}{2} = \frac{1}{\pi} \cosh^{-1}\left(1 + 2\, e^{-D_0}\right)\,.
\eeq 

In the case of solitons with different initial amplitudes,
$\dot{\Psi}_0 \neq 0$ and no initial relative phase, $\Psi_0=0$,
one finds that the solitons form an oscillatory bound state, with
their distance oscillating periodically between a minimum and a
maximum distance. Qualitative results can also be obtained for this
case, and may be found in \refref{hasegawa1995}.

\section{\label{s:numerics} Numerical simulations}

This section compares numerical simulations of \swave\
interactions of  the three different levels of description: the cold
fluid model \refeq{eq:fluid}, the \enveq\ \refeq{eq:env}, and the
\nlse\ \refneq{eq:NLS}. Where appropriate, comparisons of the
simulation results and the quasiparticle predictions of
\refsect{s:quasiparticle} are also presented.

We investigate the interactions of standing \swaves\ first found by Esirkepov\etal\rf{esirkepov1998}.
Labeling each \swave\ by an index $(j)$, we have for the cold fluid model variables:
\begin{widetext}
\begin{subequations}\label{eq:Esirkepov}
  \begin{equation}\label{eq:EsirkepovA}
  A_y^{(j)}(x,t)=
  \frac{2\sqrt{1-\omega_j^2}\cosh^2(\zeta_j)}
          {\cosh^2(\zeta_j)+\omega_j^2-1} \cos(\omega_j\,t+\theta_j)\,,\qquad
  A_z^{(j)}(x,t)=
  \frac{2\sqrt{1-\omega_j^2}\cosh^2(\zeta_j)}
          {\cosh^2(\zeta_j)+\omega_j^2-1} \sin(\omega_j\,t+\theta_j) \,,
  \end{equation}
  \begin{equation}
    E_x^{(j)} (x,t) = \frac{4\,(1-\omega_j^2)^{3/2}\,\cosh(\zeta_j)\sinh(\zeta_j)}{\left(\cosh^2(\zeta_j)+\omega_j^2-1\right)^2}\,,\qquad
    E_y^{(j)} (x,t) = \omega_j\, A_z^{(j)}\,,\quad E_z^{(j)} = -\omega_j\, A_y^{(j)}\,,
  \end{equation}
  \beq
    N^{(j)}(x,t) = 1 + \left(1-\omega_j^2\right)^2\,\frac{\cosh(4\zeta_j)-2\,(2\,\omega_j^2-1)\,\cosh(2\,\zeta_j)-3}{\left(\cosh^2(\zeta_j)-1+\omega_j^2\right)^3}\,,\quad P_x =0\,,
  \eeq
\end{subequations}
\end{widetext}
where $\zeta_j=\sqrt{1-\omega_j^2}(x-x_j)$ and $\theta_j$ is an
initial phase. The soliton amplitude is
$R_0^{(j)}\equiv|\mathbf{A_{\perp}}(x_j)|=2\sqrt{1-\omega^2}/\omega^2$.
It takes its maximum value $R_{0,\mathrm{max}}=\sqrt{3}$
for $\omega_{\mathrm{min}}=\sqrt{2/3}$, where the branch of
Esirkepov \swaves\ terminates because the minimum
local density vanishes.

In the following, we study interactions of pairs of \swaves\
with frequences $\omega_1$, $\omega_2$ and initial phase difference
$\Psi_0=\theta_2-\theta_1$, \ie, with initial conditions
given by $A_y(x,0)=A_y^{(1)}(x,0)+A_y^{(2)}(x,0)$, \etc\ For the
density we take care that the boundary condition
$N(x,0)\rightarrow1$ as $x\rightarrow\pm\infty$ is satisfied by
using the initial condition $N(x,0)=N^{(1)}(x,0)+N^{(2)}(x,0)-1$.

For the simulations using \enveq\ \refeq{eq:env}, we introduce a parameter $k_1^{(j)}$ for each soliton. We then set
$|k_1^{(1)}|=1$, which fixes $\epsilon=\sqrt{1-\omega_1^2}$ through \refeq{eq:omegaepsk1}, and introduce
the \swave\ amplitude as $\alpha_j=\sqrt{1-\omega_j^2}/\epsilon$, effectively using the freedom to choose $k_1^{(j)}=\alpha_j$.
Then, according to \refeqset{eq:arai0}{eq:aprapi} of \refappe{s:ic}, initial conditions for the \swave\ centered at $x_j=X_j/\epsilon$ are given as
\begin{subequations}\label{eq:envInit}
\beq\label{eq:aInit}
  a^{(j)}(X,0)=
  \frac{\alpha_j\cosh^2(\alpha_j(X-X_j))}
          {\cosh^2\left[\alpha_j(X-X_j)\right]+\omega_j^2-1}\, e^{-i\theta_j}\,,
\eeq
\beq\label{eq:NLSinit}
  \left.\frac{\partial a^{(j)}}{\partial T}\right|_{T=0} = \frac{i\,\alpha_j^2}{2}\,a^{(j)}(X,0)\,.
\eeq
\end{subequations}

Finally, for the simulations with the \nlse\ \refeq{eq:NLSsimple}, we expand \refeq{eq:aInit} to lowest order in $\epsilon$,
to obtain the \nlse\ soliton
\beq
  a^{(j)}(X,0)= \alpha_j\sech\left[\alpha_j(X-X_j)\right]\, e^{-i\theta_j}\,.
\eeq

All simulations are carried out with the package \texttt{XMDS2}\rf{dennis2013},
using the pseudospectral method with Fourier space evaluation of partial derivatives
and a fourth order Adaptive Runge-Kutta-Fehlberg scheme (known as ARK45) for time-stepping. 
More details on the implementation are provided in \refappe{s:numfluid}.
We validated our fluid code by verifying that a single \swave\ with $\omega_1=0.98$, \ie, of the largest amplitude $R_0^{(j)}=0.414$
studied here, follows \refeq{eq:Esirkepov} up to $t=2\times10^{4}$.

\begin{figure}[htb!]
 \begin{center}
  \includegraphics[width=\cwidth]{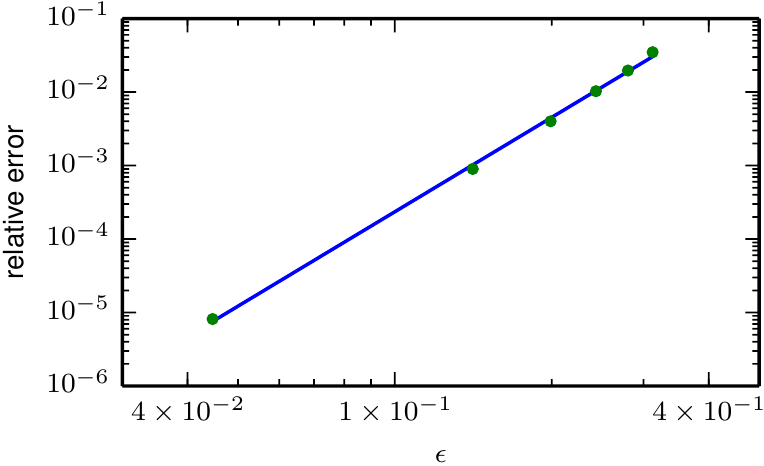}
 \end{center}
 \caption{\conline\label{f:publ_env_1s_err} Estimate of the relative error introduced by neglecting higher order terms in \enveq~\refeq{eq:env}.
  We plot $\mathrm{max}\left[a(x,T)-a(x,0)\right]/a(0,0)$ for solutions
  of \refeq{eq:env} up to $T/\epsilon^2=2\times10^4$, with initial condition \refeq{eq:envInit} with $x_1=0$.
  Dots correspond to $\omega=(0.999,\, 0.99,\, 0.98,\, 0.97,\, 0.96,\, 0.95)$.
  The solid line is the best fit to the data, showing that $\mathrm{rel.\ error}\sim \epsilon^{4.3}$.
  }
\end{figure}

Initial conditions \refeq{eq:envInit} do not correspond to an exact \swave\ solution of
the \enveq. {However, by integrating such initial conditions up to $t=T/\epsilon^2=2\times10^4$,
we show in \reffig{f:publ_env_1s_err} that the relative error introduced by the choice of initial conditions,
remains small for the maximum amplitude considered in the following numerical examples (corresponding to $\omega\geq0.98$)}.

\subsection{Small amplitude limit}

For small amplitude \swaves\ we find excellent agreement between cubic \nlse, \enveq\
and cold-fluid model predictions.

\subsubsection{Solitary waves of equal amplitude and no phase difference}

\begin{figure*}[htb!]
 \includegraphics[width=\textwidth, clip=true]{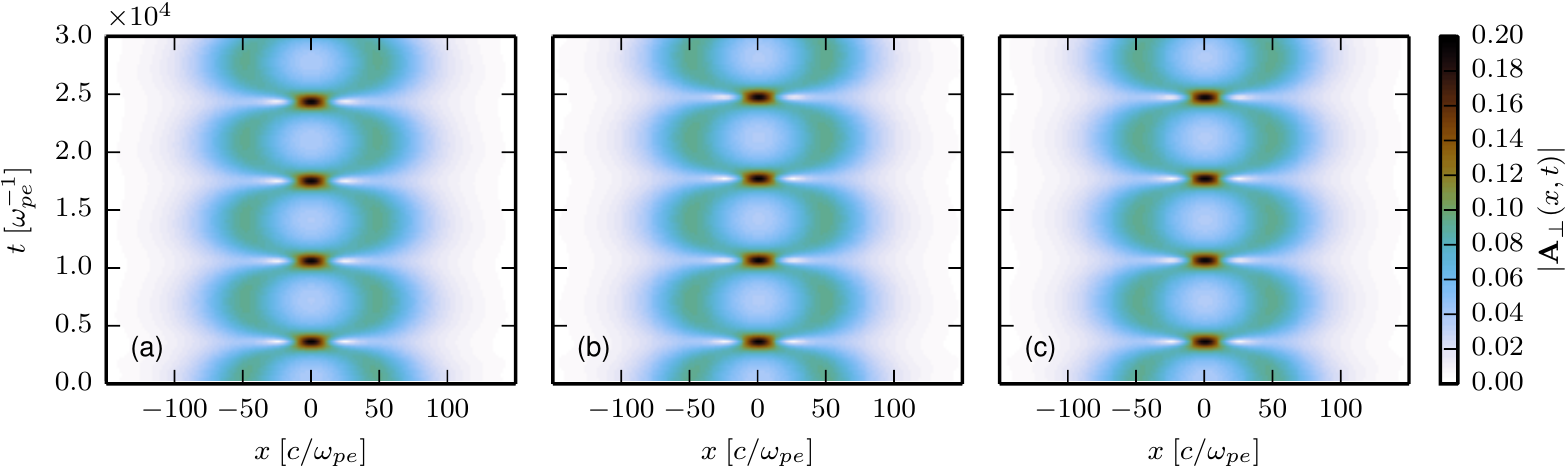}
\caption{\conline\label{f:0999_d100} Comparison of simulations of two \swave\ interaction with
frequencies $\omega_1=\omega_2=0.999$, $R_0^{(1)}=R_0^{(2)}\simeq0.0896$
(corresponding to $\epsilon\simeq0.0447$, $k_1^{(1)}=k_1^{(2)}=1$) and initial distance $d_0=100$
using (a) the fluid model \refeq{eq:fluid}, (b) \enveq\ \refeq{eq:env},
(c) \nlse\ \refneq{eq:NLS}.}
\end{figure*}

A typical case in which we have formation of a bound state of
\swaves\ is shown in \reffig{f:0999_d100}. In these
simulations, both \swaves\ have frequencies
$\omega_1=\omega_2=0.999$, $R_0^{(1)}=R_0^{(2)}\simeq0.0896$ 
(corresponding to $\epsilon\simeq0.0447$, $k_1^{(1)}=k_1^{(2)}=1$),
initial distance is $d_0=100$ and there is no initial phase
difference. The behavior of the exact cold fluid model solutions is
captured correctly by both the \nlse\ and \enveq\ models. Therefore,
at very small amplitudes, the behavior may be completely understood
in terms of the \nlse. The major role in soliton attraction and in
bound state formation is thus played by the leading term of the
relativistic nonlinearity, \ie, by the cubic term in the \nlse.

\begin{figure}[htb!]
 \begin{center}
  \includegraphics[width=\cwidth]{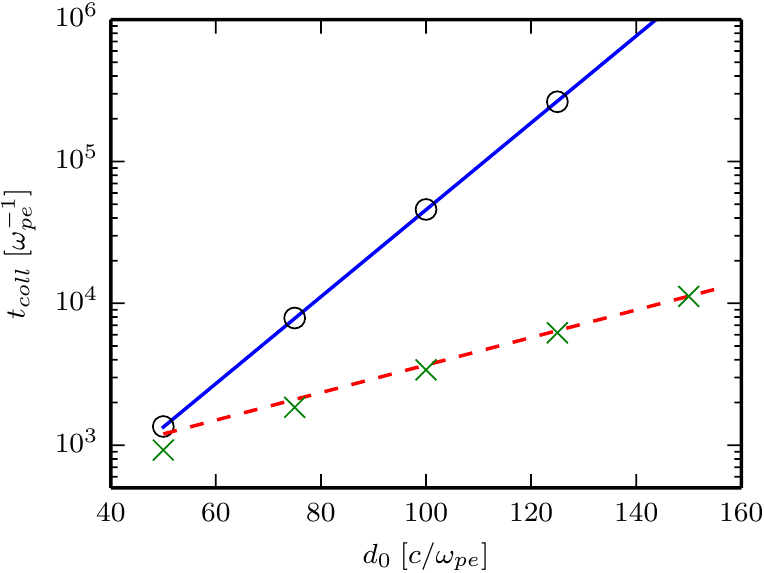}
 \end{center}
 \caption{\conline\label{f:Tc} Comparison of results of quasiparticle theory for
  $\omega=0.999$ (red, dashed line) and $\omega=0.99$ (blue, solid line)
  with numerical results from simulations of the fluid model (circles and
  cross signs, respectively)
  for the first collision time $t_{\mathrm{coll}}$ for  \swaves\ of equal amplitude
  and no initial phase difference, as a function of initial separation
  $d_0$.}
\end{figure}

The predictions of quasiparticle theory of \refsect{s:quasiparticle}
for the collision time $t_{\mathrm{coll}}$ or the period of oscillations
$t_p=2\,t_{\mathrm{coll}}$ are in excellent agreement with fluid model
simulations, see \reffig{f:Tc}. 

\subsubsection{Solitary waves of equal amplitude and finite phase difference}

\begin{figure*}[htb!]
  \includegraphics[width=\textwidth, clip=true]{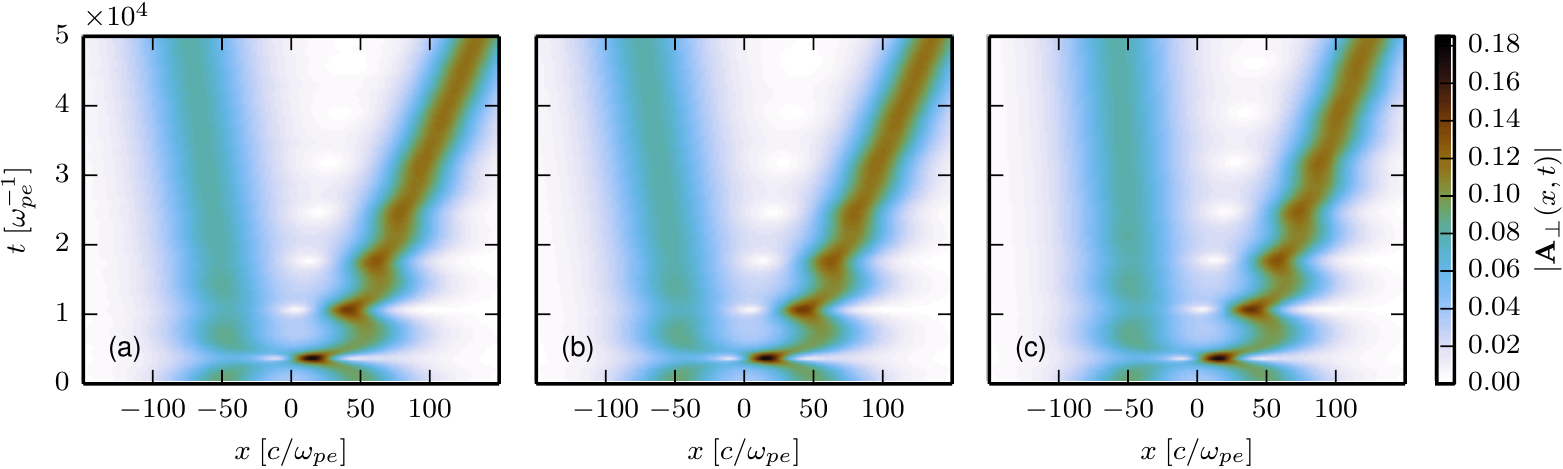}
\caption{\conline\label{f:0999_dth01} Comparison of simulations of two
\swave\ interaction with frequency $\omega_1=\omega_2=0.999$,
amplitude $R_0^{(1)}=R_0^{(2)} \simeq0.0896$ (corresponding to $\epsilon\simeq0.0447$, $k_1^{(1)}=k_1^{(2)}=1$) 
initial distance $d_0=100$ and phase difference $\Psi_0=0.1\pi$ using (a)
the fluid model \refeq{eq:fluid}, (b) \enveq\ \refeq{eq:env}, (c)
\nlse\ \refneq{eq:NLS}.}
\end{figure*}

Next, we study the case of two \swaves\ of equal frequency
$\omega_1=\omega_2=0.999$ and amplitude $R_0^{(1)}=R_0^{(2)} \simeq
0.0896$ (corresponding to $\epsilon\simeq0.0447$, $k_1^{(1)}=k_1^{(2)}=1$) 
and a finite initial phase difference
{$\Psi_0=0.1\pi$}, see \reffig{f:0999_dth01}. Also in this
case, we find excellent quantitative agreement between cold fluid
model, \nlse\ and \enveq\ simulations. The \swaves\ attract,
but do not collide, in agreement with quasiparticle theory which
predicts that collisions are inhibited for {$\Psi_0 > \Psi_c=0.136$}. 
The \swaves\ reach a minimum distance
$d_{\mathrm{min}}\simeq57$ at time $t_{\mathrm{min}} \simeq 3485$ and subsequently
drift apart. Quasiparticle theory underestimates the minimum
distance $d_{\mathrm{min}}=38.2$, while the time of minimum approach
$t_{\mathrm{min}}=3720$ is in good agreement with simulations.
\edit{According to quasiparticle theory, \refeq{eq:Da}, the solitons differ
in amplitude by $|\Delta A|= 2\,\epsilon\, |\Delta a| = 0.037$. 
This is in excellent agreement with the \nls\ simulation, for which the soliton amplitudes
differ at $t=10^5$ by $|\Delta A| = 0.036$.
}  

\subsubsection{Solitary waves of unequal amplitude and no phase difference}

\begin{figure*}[htb!]
  \includegraphics[width=\textwidth, clip=true]{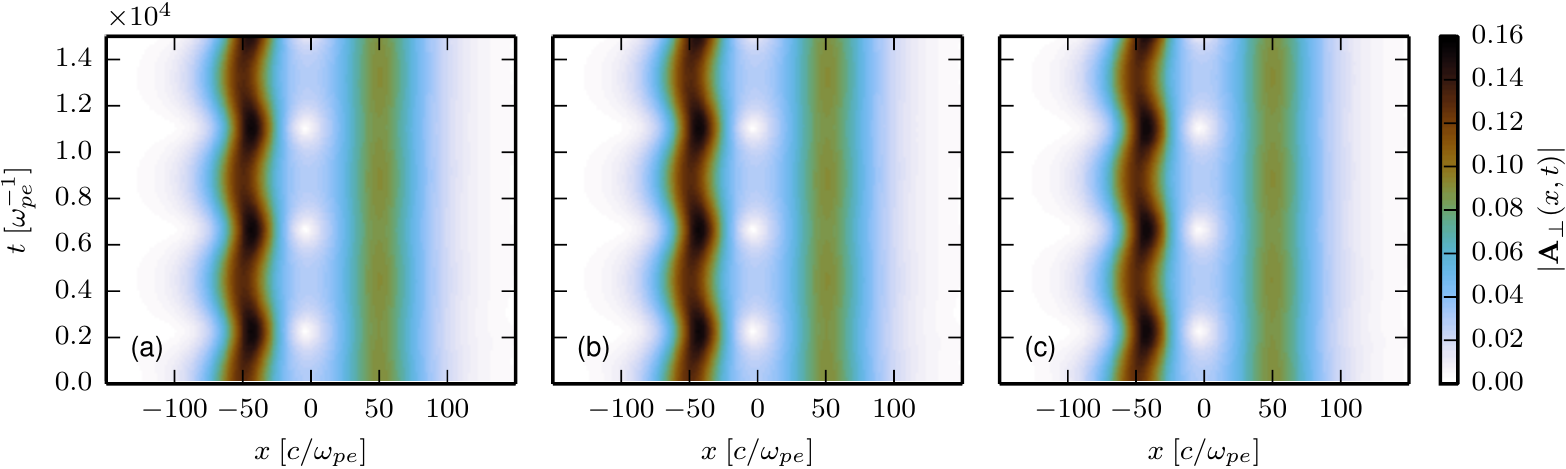}
\caption{\conline\label{f:0999_0998_d100} Comparison of simulations of two \swave\ interaction with
$\omega_1=0.999$, $R_0^{(1)}\simeq0.0896$  
and $\omega_2=0.998$, $R_0^{(2)}\simeq0.127$ (corresponding to $\epsilon\simeq0.0447$, $k_1^{(1)}=1$, $k_1^{(2)}\simeq1.4139$) 
and initial distance $d_0=100$
using (a) the fluid model \refeq{eq:fluid}, (b) \enveq\ \refeq{eq:env},
(c) \nlse\ \refneq{eq:NLS}.}
\end{figure*}

For \swaves\ of unequal frequences $\omega_1=0.999$ and $\omega_2=0.998$,
and therefore also unequal amplitudes $R_0^{(1)}\simeq0.0896$, 
$R_0^{(2)}\simeq0.127$ (corresponding to $\epsilon\simeq0.0447$, $k_1^{(1)}=1$, $k_1^{(2)}\simeq1.4139$),
with no initial phase difference, as shown for example in \reffig{f:0999_0998_d100},
the \swaves\ interact and form a periodic bound state.
However, their separation remains finite. Again, there is
excellent agreement between all three levels of description.
This behavior can be also understood in terms of
quasiparticle theory of \nlse, see \refref{hasegawa1995}.

\subsection{\label{s:intense} Larger amplitudes}

For \swaves\ of moderately large amplitude, but still in the perturbative regime $\epsilon\ll1$,
cold-fluid model simulations deviate from cubic \nlse\ predictions
and the fifth order terms need to be taken into account.

\subsubsection{Solitary waves of equal amplitude and no phase difference}

\begin{figure*}[ht!]
 \includegraphics[width=\textwidth, clip=true]{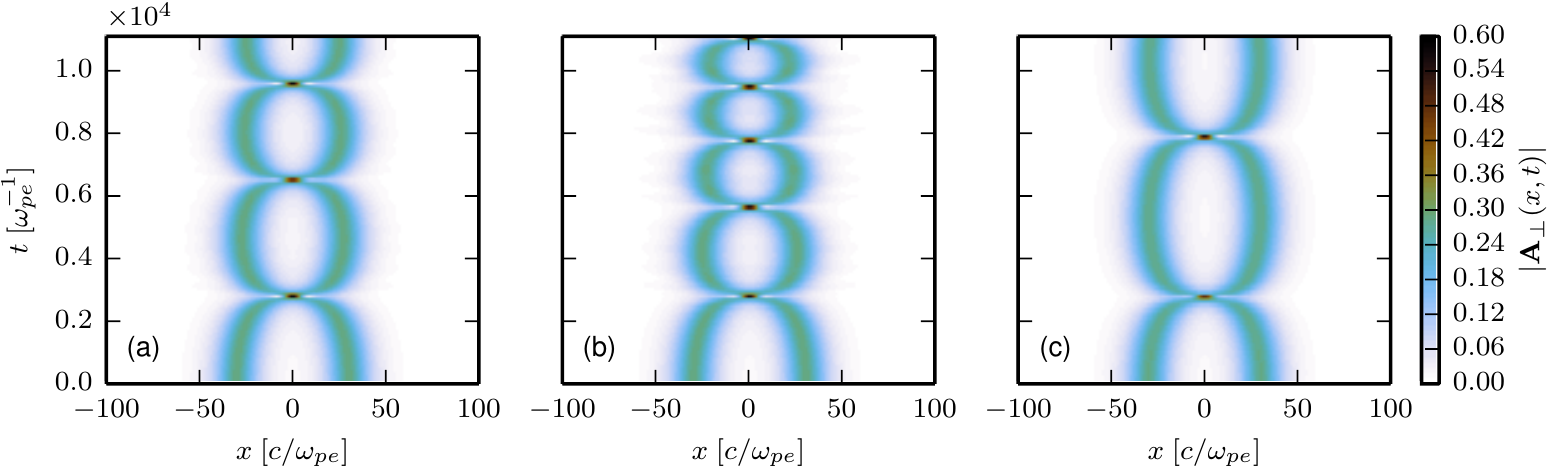}
\caption{\conline\label{f:099_d60} Comparison of simulations of \swave\ interaction with
frequencies $\omega_1=\omega_2=0.99$, amplitude $R_0^{(1)}=R_0^{(2)} \simeq0.2879$ ($\epsilon\simeq0.1411$, $k_1^{(1)}=k_1^{(2)}=1$)
and initial distance $d_0=60$ using
(a) the fluid model \refeq{eq:fluid}, (b) \enveq\ \refeq{eq:env},
(c) \nlse\ \refneq{eq:NLS}.}
\end{figure*}

\begin{figure}[ht!]
 \includegraphics[width=\cwidth, clip=true]{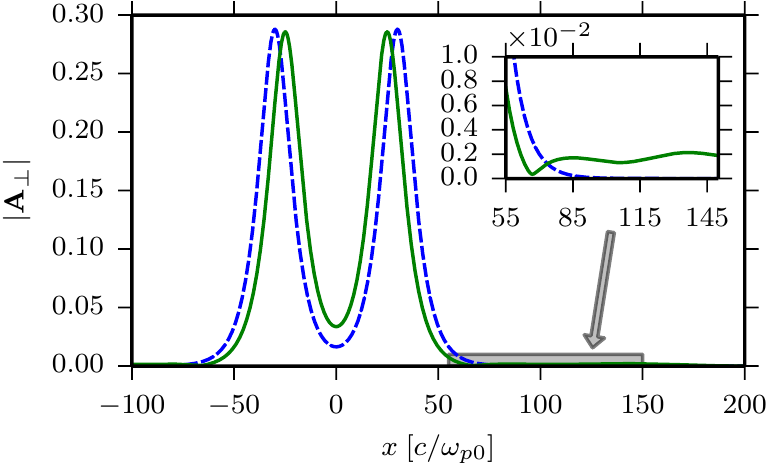} 
\caption{\conline\label{f:099_d60_profiles} Two snapshots  from cold fluid model simulations of \swave\ interaction
with
frequencies $\omega_1=\omega_2=0.99$, amplitude $R_0^{(1)}=R_0^{(2)} \simeq0.2879$ 
($\epsilon\simeq0.1411$, $k_1^{(1)}=k_1^{(2)}=1$) and initial distance $d_0=60$
[see also \reffig{f:099_d60}(a)].
Dashed (blue) line corresponds to $t=0$; solid (green) line
corresponds to $t\simeq10745$, \ie, at maximum separation after the third collision.
\edit{The inset corresponds to the area in the gray box.}
}
\end{figure}

\refFig{f:099_d60} shows the interaction of two \swaves\ of equal amplitude
with $\omega_1=\omega_2=0.99$ ($R_{0,1}=R_{0,2}\simeq0.2879$, $\epsilon\simeq0.1411$, $k_1^{(1)}=k_1^{(2)}=1$),
separated by $d_0=60$ and no initial phase difference.
The \swaves\ collide approximately at $t_{\mathrm{coll}}\simeq2800$, in good
agreement with quasiparticle theory prediction $t_{\mathrm{coll}}\simeq2717$,
and they form a bound state.
However, contrary to the prediction of the cubic \nlse\ model,
collisions are inelastic and part of the \swave\ field decays away
each time the \swaves\ collide, see \reffig{f:099_d60_profiles}. The bound state
is reminiscent of a system of damped oscillators: after each encounter,
the \swaves\ decrease in amplitude and come closer together. In turn,
this implies that the period of bound state oscillations becomes shorter,
as predicted by \refeq{eq:tc}.

This behavior is also captured qualitatively by the \enveq, see \reffig{f:099_d60}(b).
However, the \enveq\ overestimates the loss of \swave\ field at collisions and,
correspondingly, the period of oscillations decreases with a faster rate than
in the fluid model simulations. These deviations may be attributed to the
breakdown, at the moment of collision,
of the assumptions of small field magnitude and slow spatio-temporal evolution.
In particular, close to the collision, the spatial
scale for variations in the envelope is a few $c/\omega_{pe}$,
indicating that we should expect quantitative
discrepancies between the fluid and \enveq\ simulations.

The cubic \nlse\ approximation on the other hand, presents qualitative difference in
this case. Since cubic \nlse\ is an integrable model, all soliton collisions are
elastic, leading to an exactly periodic bound state of \swaves.
Therefore, the inclusion of the fifth order terms in our analysis is necessary
for a qualitatively correct description of \swave\ interactions.

The collision time on the other hand, as predicted by quasiparticle theory of \nlse,
agrees very well with fluid simulation results even in this larger amplitude case,
see \reffig{f:Tc}. The reason for this is that attraction of \swaves\ is determined by
their overlap. For large initial separation of the
\swaves\, this overlap occurs at small amplitudes. Moreover, in the limit
$x\rightarrow\pm\infty$ the exact \swave\ envelope \refeq{eq:aInit} takes the form of
the \nlse\ soliton \refeq{eq:NLSinit}. Therefore, \nlse\ approximation is a valid one
in order to determine the initial attraction phase of two well separated \swaves\,
even for larger amplitudes.

\subsubsection{Solitary waves of equal amplitude and finite phase difference}

\begin{figure*}[htb!]
  \includegraphics[width=\textwidth, clip=true]{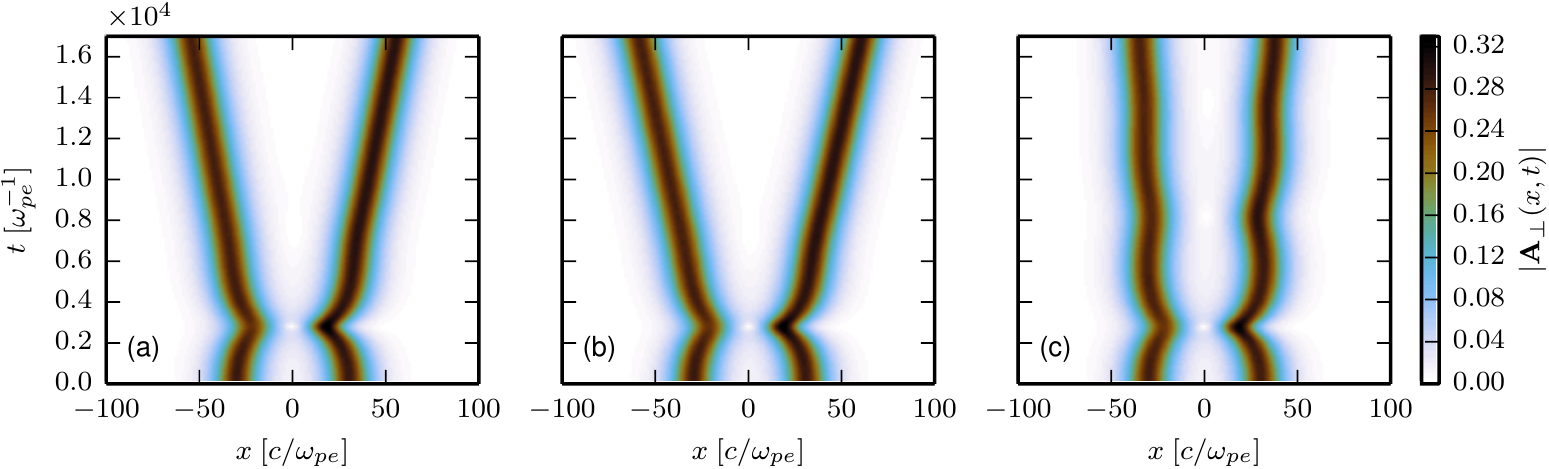}
\caption{\conline\label{f:099_dth01} Comparison of simulations of two
\swave\ interaction with frequency $\omega_1=\omega_2=0.99$
(amplitude $R_0^{(1)}=R_0^{(2)} \simeq0.2879$, $\epsilon\simeq0.1411$, $k_1^{(1)}=k_1^{(2)}=1$) 
initial distance $d_0=60$ and phase difference $\Psi_0=0.1\pi$ using (a) the
fluid model \refeq{eq:fluid}, (b) \enveq\ \refeq{eq:env}, (c) \nlse\
\refneq{eq:NLS}.}
\end{figure*}

As an example, we show the interaction of two \swaves\ with
$\omega_1=\omega_2=0.99$ ($R_{0,1}=R_{0,2}\simeq0.2879$, $\epsilon\simeq0.1411$, $k_1^{(1)}=k_1^{(2)}=1$) 
with initial distance $d_0=60$ and phase difference $\Psi_0=0.1\,\pi$,
in \reffig{f:099_dth01}. The fluid simulations show that the
\swaves\ initially approach and then separate, moving in
opposite directions with velocities $v_L=-0.002$ and $v_R=0.0021$
for the left and right moving soliton, respectively. The \enveq\
simulations agree with the fluid model predictions quantitatively,
faithfully capturing the velocities of the outgoing \swaves\
to be $v_L=-0.0024$ and $v_R=0.0025$. On the other hand, \nlse\
simulations exhibit oscillations in soliton position before the
latter drift apart, a feature not present in the fluid simulations.
Furthermore, the velocities of the outgoing \swaves\ in the
\nlse\ simulations are $v_L=-0.0005$ and $v_R=0.0007$, rather small
compared to the fluid simulations.

Quasiparticle predictions for the distance of minimum approach
$d_{\mathrm{min}}$ and corresponding time $t_{\mathrm{min}}$ are compared with the results of
the numerical simulations in \reftab{tab:mindist}, showing excellent agreement.

\begin{table}[ht!]
  \begin{tabular}{l|c|c}
                & $d_{\mathrm{min}}\, [c/\omega_{pe}]$   & $t_{\mathrm{min}}$ $[\omega_{pe}^{-1}]$  \\ \hline
  quasiparticle (analytical)    & 40.5                          & 2751.4 \\
  fluid simulation      & 39.8                          & 2703.0 \\
  \enveq\ simulation        & 39.7                          & 2682.2 \\
  \nlse\ simulation      & 40.0                          & 2688.4
  \end{tabular}
 \caption{\label{tab:mindist} Comparison of analytical
  prediction based on quasiparticle theory and results of numerical simulations
  of the different models studied here, for the distance of minimum approach
  $d_{\mathrm{min}}$ and corresponding time $t_{\mathrm{min}}$, for two \swaves\ of frequency $\omega_1=\omega_2=0.99$,
  amplitude $R_0^{(1)}=R_0^{(2)} \simeq0.2879$ ($\epsilon\simeq0.1411$, $k_1^{(1)}=k_1^{(2)}=1$), 
  initial distance $d_0=60$ and initial phase difference
  $\Psi_0=0.1\pi$
  (see \reffig{f:099_dth01}).
  }
\end{table}

\begin{figure*}[htb!]
 \includegraphics[width=\textwidth, clip=true]{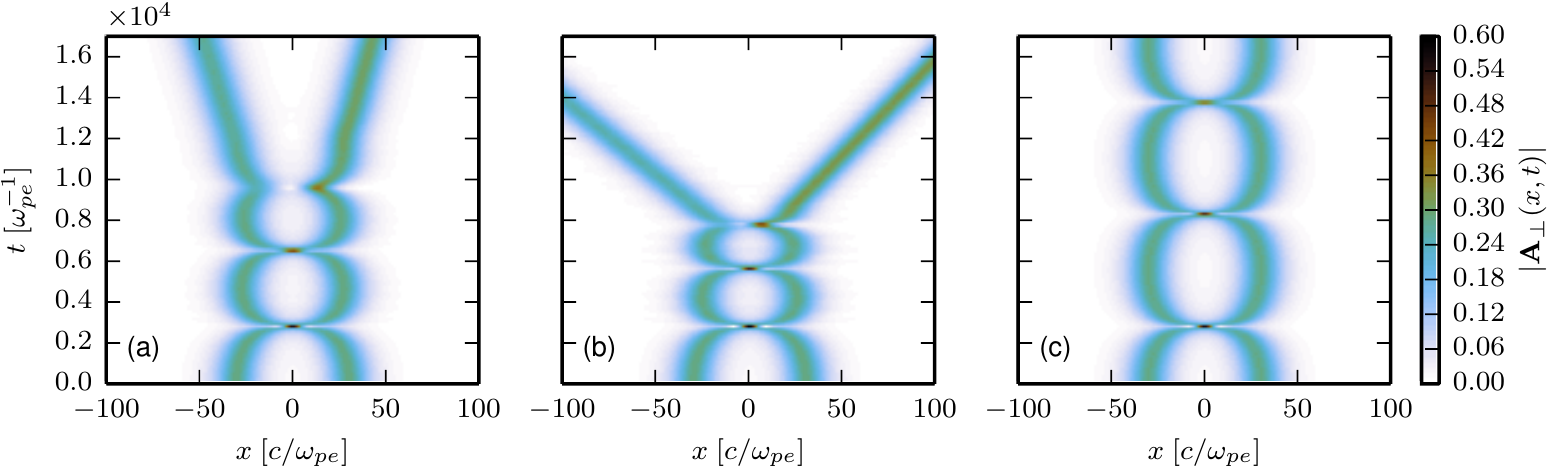}
\caption{\conline\label{f:099_dth1e-6} Comparison of simulations of two
\swave\ interaction with frequency $\omega_1=\omega_2=0.99$,
amplitude $R_0^{(1)}=R_0^{(2)} \simeq0.2879$ ($\epsilon\simeq0.1411$, $k_1^{(1)}=k_1^{(2)}=1$)  initial distance
$d_0=60$ and phase difference {$\Psi_0=10^{-6}$} using (a)
the fluid model \refeq{eq:fluid}, (b) \enveq\ \refeq{eq:env}, (c)
\nlse\ \refneq{eq:NLS}.}
\end{figure*}

We also present the interaction of two \swaves\ with
$\omega_1=\omega_2=0.99$ ($R_{0,1}=R_{0,2}\simeq0.2879$, $\epsilon\simeq0.1411$, $k_1^{(1)}=k_1^{(2)}=1$), 
initial distance $d_0=60$ and very small phase difference
$\Psi_0=10^{-6}$ in \reffig{f:099_dth1e-6}. In this case,
$\Psi_0<\Psi_c$, and indeed the two \swaves\ do
collide. However, after a subsequent recollision, they diverge and
move away from each other. The \enveq\ simulations capture this
feature, even though the velocities of the escaping \swaves\
are much larger than in the fluid simulations. However, in the
\nlse\ simulations we see that the solitons keep recolliding for
many iterations. \edit{Quasiparticle theory predicts that even for small
initial phase difference the solitons will eventually drift apart; however,
from \refeq{eq:DwPhase} we find that the time scale required for this to
happen for such a small initial phase difference is 
of the order of $10^{9}$, well beyond the maximum integration time $t=2\times10^5$ 
for which we could simulate the system.}  

\begin{figure}[ht!]
 \includegraphics[width=\cwidth, clip=true]{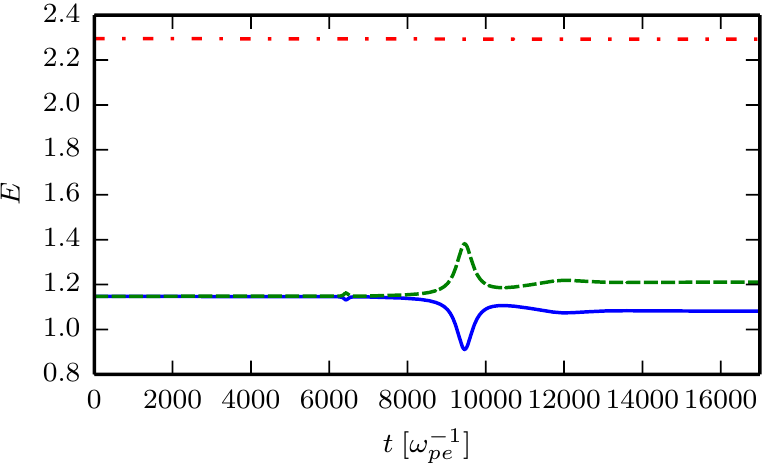} 
\caption{\conline\label{f:energyExchange} 
Total energy $E_{\mathrm{tot}}$ in the computational domain (red, dot-dashed line)
and energy $E_L$ (blue, solid line) and $E_R$ (green, dashed line) in the left and right half of the domain,
respectively, corresponding to cold fluid model simulation of \reffig{f:099_dth1e-6}(a).
}
\end{figure}

\edit{The cold fluid model conserves the normalized energy, $E=E_l+E_p+E_e$,
where
\begin{align}
E_l &= \frac{1}{2}\int \left[\left(\frac{\partial A_y}{\partial
t}\right)^2+\left(\frac{\partial A_y}{\partial
x}\right)^2\right]dx\,,\\
E_p &= \frac{1}{2}\int \left(\frac{\partial \phi}{\partial
x}\right)^2dx\,,\\
 E_e &= \int \left(\gamma-1\right)n\ dx\,,
\end{align}
are electromagnetic, electrostatic and kinetic energy contributions, 
respectively. In soliton collisions with a finite phase difference, energy
can be transfered between the two solitons. We show this in \reffig{f:energyExchange}, 
where we plot as a function of time the total energy $E_{\mathrm{tot}}$ in the computational domain
and the energy $E_L$ ($E_R$) in the left (right) half of the domain, for the 
cold fluid model simulation of \reffig{f:099_dth1e-6}(a).
We find that after the solitons separate, energy has been transfered to the 
soliton in the right half of the domain, while its amplitude has 
increased.}

\begin{figure*}[ht!]
 \includegraphics[width=\textwidth, clip=true]{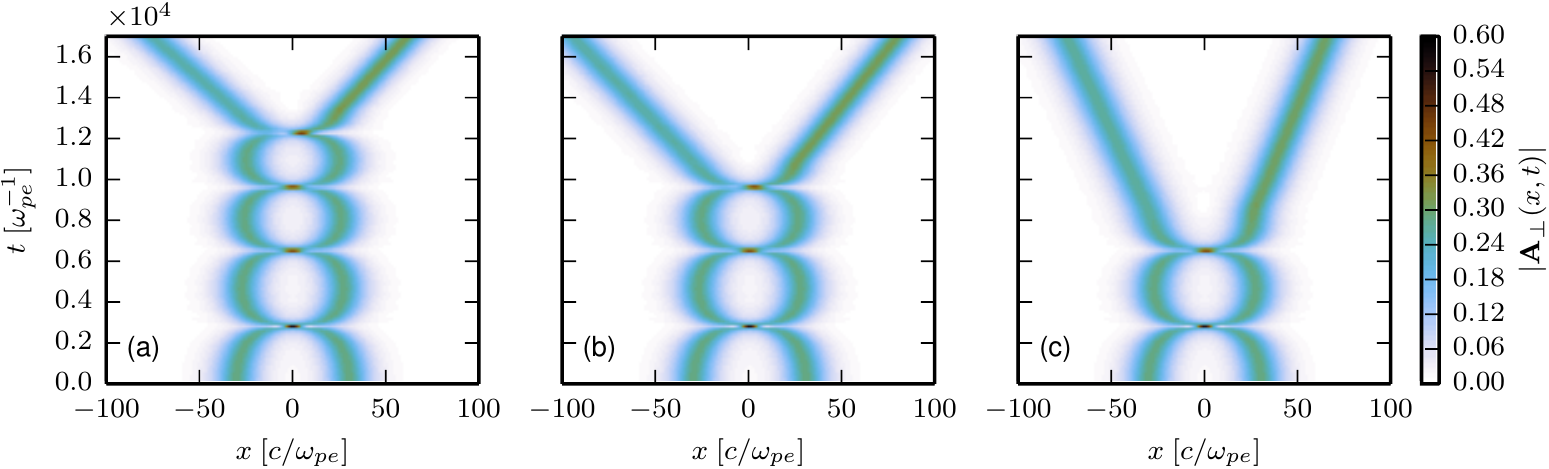}
\caption{\conline\label{f:099_dth} Cold fluid model simulations of two
\swave\ interaction with frequency $\omega_1=\omega_2=0.99$,
amplitude $R_0^{(1)}=R_0^{(2)} \simeq0.2879$ ($\epsilon\simeq0.1411$, $k_1^{(1)}=k_1^{(2)}=1$)  initial distance
$d_0=60$ and phase difference using (a) {$\Psi_0=10^{-9}$}
(b) {$\Psi_0=10^{-7}$}, (c) {$\Psi_0=10^{-5}$}.} 
\end{figure*}

\edit{The results of \reffig{f:099_dth1e-6} indicate 
that for larger amplitudes, cold fluid model bound states 
do not persist under perturbations involving phase difference of the
two \swaves, leading to qualitative difference from the
\nlse\ model dynamics. \reffig{f:099_dth} illustrates that 
the number of re-collisions between solitons decreases as $\Psi_0$
increases. Such dynamics has been associated with chaotic scattering
of solitons in the context of a perturbed NLS equation\rf{dmitriev2002}. 
However, determining parameters for which chaotic scattering occurs
in our problem is beyond the scope of this work. 
\reffig{f:099_dth} suggests that truncation error in simulations 
can lead to the breaking of a bound state, even when there is 
no initial phase difference, as we have observed in
\refref{saxena2013}. Therefore, we have taken care that simulations
presented here are well resolved by checking that varying the spatial resolution
does not affect the results.
} 

\subsubsection{Solitary waves of unequal amplitude and no phase difference}

\begin{figure*}[htb!]
  \includegraphics[width=\textwidth, clip=true]{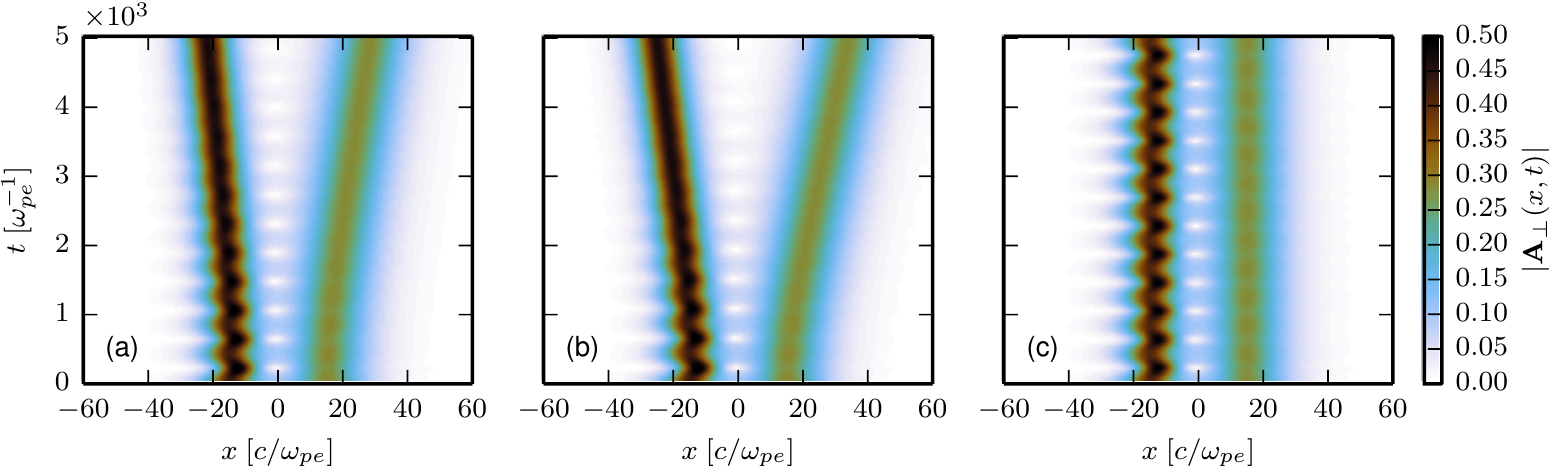}
\caption{\conline\label{f:099_098_d30} Comparison of simulations of two \swave\ interaction with
frequencies  $\omega_1=0.99$, 
$\omega_2=0.98$, amplitudes $R_0^{(1)}\simeq0.2879$, $R_0^{(2)}\simeq 0.4144$
(corresponding to $\epsilon\simeq0.1411$, $k_1^{(1)}=1$, $k_1^{(2)}\simeq1.4107$), 
initial distance $d_0=30$ and relative phase
{$\Psi_0=0$}
using (a) the fluid model \refeq{eq:fluid}, (b) \enveq\ \refeq{eq:env},
(c) \nlse\ \refneq{eq:NLS}.}
\end{figure*}

The case of \swaves\ of frequencies $\omega_1=0.99$, 
$\omega_2=0.98$, amplitudes $R_0^{(1)}\simeq0.2879$, $R_0^{(2)}\simeq 0.4144$
(corresponding to $\epsilon\simeq0.1411$, $k_1^{(1)}=1$, $k_1^{(2)}\simeq1.4107$), 
initial distance $d_0=30$ and relative phase
{$\Psi_0=0$}, is shown in \reffig{f:099_098_d30}. In this
case, the \swaves\ after some oscillations quickly diverge
from each other. Simulations using the \enveq\ faithfully capture
this behavior. On the contrary, \nlse\ simulations show the
formation of a bound state of oscillating solitons. We can therefore
conclude that once again we have qualitative differences between
cold-fluid model and \nlse\ \swave\ interactions, with bound
states appearing unstable within the former model. 

\subsection{\label{s:cqnlse} Cubic-quintic \nlse\ approximation}

Finally, we show that the qualitatively new features present in the larger amplitude
simulations, may be captured by keeping only the quintic nonlinearity in \refeq{eq:env},
\ie, by the cubic-quintic \nlse:
\beq\label{eq:cqnlse}
    i\frac{\partial a}{\partial T}+\frac{1}{2}\frac{\partial^2 a}{\partial X^2}+|a|^2 a - 3\,\epsilon^2 \, |a|^4 a = 0\,,
\eeq

\begin{figure*}[hbt!]
  \includegraphics[width=\textwidth, clip=true]{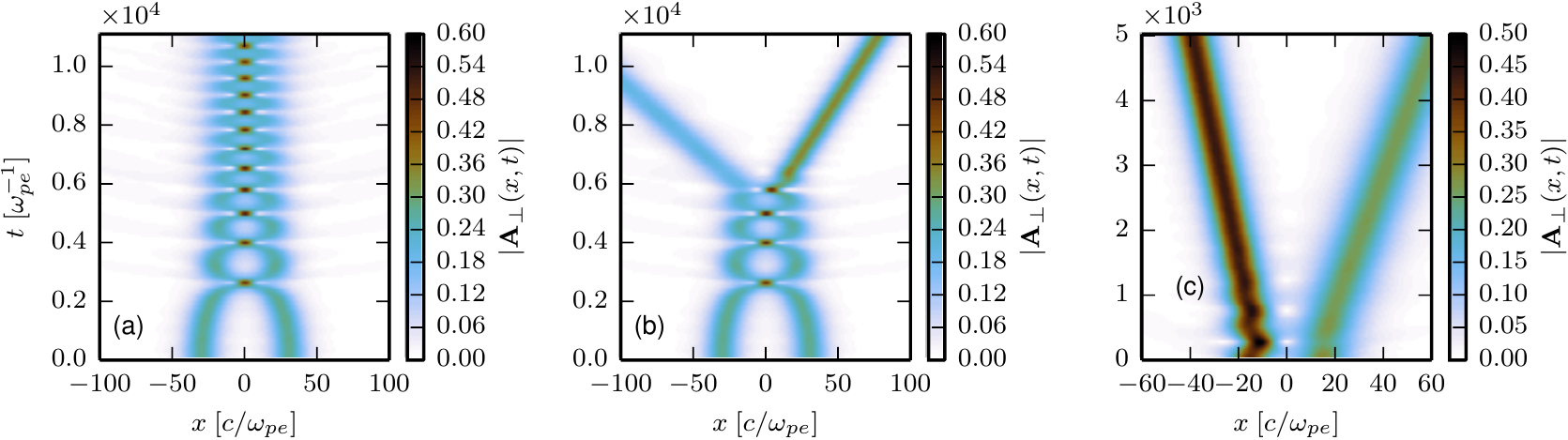}
\caption{\conline\label{f:cqnlse} Solitary wave interaction using the
cubic-quintic \nlse\ approximation \refeq{eq:cqnlse} for two
\swaves\ with (a) frequencies $\omega_1=\omega_2=0.99$,
amplitudes $R_0^{(1)}=R_0^{(2)} \simeq0.288$ and initial distance
$d_0=60$, as in \reffig{f:099_d60} , (b) frequencies
$\omega_1=\omega_2=0.99$, amplitudes $R_0^{(1)}=R_0^{(2)}
\simeq0.288$ initial distance $d_0=60$ and phase difference
{$\Psi_0=10^{-6}$}, as in \reffig{f:099_dth01} and (c)
frequencies $\omega_1=0.99$, $\omega_2=0.98$ ($R_0^{(1)}\simeq0.288$
and $R_0^{(2)}\simeq 0.414$) and initial distance $d_0=30$, as in
\reffig{f:099_098_d30}. }
\end{figure*}

We present three different examples of \swave\ interaction dynamics under \refeq{eq:cqnlse}
in \reffig{f:cqnlse}. In all cases, there is agreement at a qualitative level with cold fluid
model simulations of \refeq{eq:cqnlse} of \refsect{s:intense}. This suggests that the
'defocusing' quintic term induces the qualitative changes in \swave\ interactions.
However, obtaining better quantitative agreement requires keeping all terms in \refeq{eq:env},
as in \refsect{s:intense}.

\section{\label{s:concl} Discussion and Conclusions}

We studied weakly relativistic bright \swaves\ and their interactions
using a perturbative, multiple scale analysis and direct numerical simulations.
We derived an equation for the evolution of the field envelope valid for
small amplitudes, keeping terms up to order five in the small parameter.

Localization of the \swave\ solutions appears naturally in our scheme
through the requirement $\omega\lesssim\omega_{pe}$ obtained through the linear
dispersion relation. The lowest order nonlinear effect is due to the
relativistic nonlinearity, leading to a classical cubic \nlse.
The cubic nonlinearity balances pulse dispersion, leading to the
formation of \swaves. However, higher order terms, most importantly
the quintic term resulting from the expansion of the relativistic
nonlinearity, become essential at larger amplitudes. In particular,
the response of the plasma to the ponderomotive force and the formation of a
density cavitation which confines the \swave, is only captured
by keeping fifth order terms in the small parameter, leading to the
\enveq\ \refeq{eq:env}.
At even higher amplitudes, \swave\ width becomes comparable to the
wavelength \edit{of the carrier electromagnetic wave} 
and the perturbative description breaks down.

We have demonstrated the utility of the \nlse\ and \pNLS\ equations
derived here by applying them to the problem of standing solitary
wave interaction. We have found that the lowest order, \nlse\
approximation works very well for lower amplitudes, but gives
qualitatively different results at higher amplitudes. For well
separated solitons the quasiparticle approach provides analytical
estimates for the first collision time and minimum distance of
approach of two solitons. We have found that these estimates are in
very good agreement with fluid simulations, even for larger
amplitudes. The reason for this is that the overlapping part of well
separated relativistic \swaves\ can be well approximated by
the tails of \nlse\ solitons. The effect of higher order terms that
leads, \eg, to inelastic collisions, only becomes significant after
the \swaves\ have approached each other.

Once the \swaves\ are sufficiently close to each other, the
higher order terms become important, and lead to qualitatively
different results than in \nlse. For example, since the \nlse\ is a
completely integrable equation, its soliton collisions are elastic.
However, in our fluid simulations we have clear signatures of
inelastic \swave\ collisions accompanied by emission of
radiation. These features are captured qualitatively by keeping the
higher order terms in the \enveq. We can attribute the
emission of radiation in collisions to the role of the
'defocusing' fifth order nonlinearity. In cases in which 
the total amplitude of the field remains in
the perturbative regime, good quantitative agreement is obtained
between the cold-fluid and \enveq\ simulations.
{Our study suggests that the collision time
for well separated \swaves\ can become larger than the typical
response time of the ions $\left(m_i/m_e\right)^{1/2}\,2\pi\,/\omega_{pe}$\rf{naumova2001},
and ion dynamics would have to be taken into account in future studies.}

\edit{As we have seen in \reffig{f:energyExchange}, one feature 
of soliton collisions is the transfer of
energy from one soliton to the other during the interaction. Moreover, it can be
seen in \reffig{f:099_dth1e-6}, that the wave on the right hand
side has a larger amplitude after the collision.} This is 
interesting in connection to wave-breaking of \swaves, which 
occurs above a certain amplitude threshold, and has been 
proposed in the past as a means to
accelerate particles \cite{esirkepov1998,Farina01a}. Our
simulations indicate that \swave\ interaction is a good
candidate to trigger wave-breaking, through which 
electromagnetic energy of the wave could be transfered to the particles.

The introduction of NLS equation as the lowest order approximation
to the problem of relativistic solitary interaction allowed the application
of quasiparticle theory of \nlse\ solitons
in order to obtain analytical estimates for collision time and minimum distance
of approach. The development of the 
\enveq\ framework, on the other hand, suggests the possibility to use further
\edit{mathematical tools such as the inverse scattering transform (IST)~\rf{hasegawa1995}.
For instance, in the
context of Alfven waves, IST applied to the derivative-\nlse\ was used to explain
the collapse of the bright Alfven soliton and the formation of
robust magnetic holes \rf{Sanchez2010a,Sanchez2010b}.
The model 
derived here suggests that the IST of the \nlse\ could be used in a similar manner to analyze
simulations and get insight on the interaction and
disappearance of \swaves\ in laser-plasma interaction.}

{In summary, a perturbed NLS equation describing electromagnetic envelope
evolution for weakly relativistic pulses in plasmas has been derived. 
Auxiliary equations describe the plasma density, momentum and electrostatic potential 
in terms of the electromagnetic field. The pNLS model agrees very well with fluid model 
simulations of \swave\ interactions. Our simulations suggest 
that in the small but finite amplitude regime the defocusing quintic nonlinearity becomes important
and \swave\ collisions are inelastic.
}

\section{Acknowledgements}

ES would like to thank S. Skupin and F. Maucher for helpful suggestions.
Work by GS was supported by the Ministerio de Ciencia e Innovaci\'{o}n of Spain (Grant No
ENE2011-28489). We would like to thank the anonymous referee for many
perspicacious observations.

\appendix

\section{\label{s:ic}Relation of fluid model and envelope initial conditions}

Here, we specify how initial conditions of the cold fluid model are related
to initial conditions of the \enveq. Letting $a = a_r+i\,a_i$, we note that
\refEqs{eq:expAy}{eq:expAz} and \refeq{eq:a0corr} with the help of $a=-i\,b$ give
\small
\begin{subequations}\label{eq:AyAzEnv}
\begin{align}
    A_y(x,t) &= 2\, \epsilon\, \left[a_r(X,T)\, \cos\, t +a_i(X,T) \sin\,t\right]\,, \\
    A_z(x,t) &= 2\, \epsilon\, \left[-a_i(X,T)\, \cos\, t + a_r(X,T) \sin\, t\right]\,,
\end{align}
\end{subequations}
\normalsize
or
\begin{subequations}\label{eq:araiXT}
\begin{align}
    a_r(X,T) &=  \frac{1}{2\,\epsilon} (A_y\, \cos\, t + A_z \sin\,t)\,, \\
    a_i(X,T) &=  \frac{1}{2\,\epsilon} ( - A_z \cos\,t + A_y \, \sin\, t)\,.
\end{align}
\end{subequations}
Setting $T=t=0$ we get
\begin{subequations}\label{eq:arai0}
\begin{align}
    a_r(X,0) &=  \frac{1}{2\,\epsilon} A_y(x,0)\,,\\
    a_i(X,0) &=  -\frac{1}{2\,\epsilon} A_z(x,0)\,.
\end{align}
\end{subequations}

Using \refeq{eq:dispAppr} in \refeq{eq:araiXT} we arrive at
\small
\begin{widetext}
\begin{subequations}\label{eq:arai}
  \begin{align}
    a_r &= \frac{1}{2\,\epsilon} \left[A_y\, \cos\,\omega\, t \cos\left(\frac{|k_1|^2}{2} T\right) - A_y\, \sin\,\omega\, t \sin\left(\frac{|k_1|^2}{2} T\right) \right.
        \left. + A_z \sin\,\omega t \cos\left(\frac{|k_1|^2 T}{2}\right) +A_z \cos\,\omega t \sin\left(\frac{|k_1|^2}{2} T\right)\right]\,,\\
    a_i &= \frac{1}{2\,\epsilon} \left[-A_z\, \cos\,\omega\, t \cos\left(\frac{|k_1|^2}{2} T\right) + A_z\, \sin\,\omega\, t \sin\left(\frac{|k_1|^2}{2} T\right) \right.
        \left. + A_y \sin\,\omega t \cos\left(\frac{|k_1|^2 T}{2}\right) +A_y \cos\,\omega t \sin\left(\frac{|k_1|^2}{2} T\right)\right]\,.
  \end{align}
\end{subequations}
\end{widetext}
\normalsize
Taking the derivative of \ref{eq:arai} with respect to $T$ and setting $t=T=0$ we get
\begin{subequations}\label{eq:aprapi}
\begin{align}
  \left.\frac{\partial a_r}{\partial T}\right|_{T=0} &= \frac{A_z(x,0)}{4\epsilon}|k_1|^2\,,\\
  \left.\frac{\partial a_i}{\partial T}\right|_{T=0} &= \frac{A_y(x,0)}{4\epsilon}|k_1|^2\,.
\end{align}
\end{subequations}

\section{\label{s:numfluid} Numerical solution of fluid equations}

For the numerical solution of the fluid system we chose to use spectral (Fourier)
discretization of the field and plasma quantities, while time stepping is
handled by an adaptive fourth order Runge-Kutta scheme.
This necessitates the introduction of the
components of the electric field, in order to arrive at partial differential
equations involving only first time derivatives. Specifically, we
write the longitudinal component of Ampere's law as
  \begin{equation}\label{eq:longitudinalNR}
    \frac{\partial E_x}{\partial t} = \frac{N}{\gamma}P_x\,,
  \end{equation}
  while the wave equation is split into
  \begin{equation}\label{eq:waveENR}
    \frac{\partial\mathbf{\Etr}}{\partial t}=-\frac{\partial^2\textbf{\Atr}}{\partial x^2}+\frac{N}{\gamma}\textbf{\Atr}\,,
  \end{equation}
  and
  \begin{equation}\label{eq:waveANR}
    \frac{\partial\mathbf{\Atr}}{\partial t}=-\mathbf{\Etr}\,.
  \end{equation}
  The momentum equation yields simply
  \begin{equation}\label{eq:PxNR}
    \frac{\partial P_x}{\partial t}=-E_x-\frac{\partial \gamma}{\partial x}\,.
  \end{equation}
  Instead of solving Poisson \refeq{eq:Poisson}, we introduce the continuity equation in order to determine the rate of change of
  density,
  \begin{equation}
    \frac{\partial N}{\partial t} = -\frac{\partial}{\partial x}\left(\frac{N\,P_x}{\gamma}\right)\,.
  \end{equation}
  Finally, using \refeq{eq:gamma} we derive an equation for the rate of change of $\gamma$,
  \begin{equation}
   \frac{\partial \gamma}{\partial t} = -\frac{1}{\gamma}\left(P_x\left(E_x+\frac{\partial \gamma}{\partial x}\right)+A_y\,E_y+A_z\,E_z\right)\,.
  \end{equation}


%


\end{document}